\documentclass[twocolumn]{aastex631}
\usepackage{appendix,natbib}
\usepackage{amsmath}
\usepackage{courier}
\usepackage{booktabs}
\usepackage{upgreek}

\newcommand{\zsys}{z$_{sys}$}
\newcommand{\mgii}{\ion{Mg}{2}}
\newcommand{\oii}{\ion{[O}{2]}}

\newcommand{\kmps}{\ensuremath{\mathrm{km~s}^{-1}}}

\begin{document}

\title{The Outflowing \oii\ Nebulae of Compact Starburst Galaxies at $z\sim 0.5$}
\correspondingauthor{Serena Perrotta}
\email{s2perrotta@ucsd.edu}

\author{Serena Perrotta}

\author{Alison L. Coil} 
\affiliation{Department of Astronomy and Astrophysics, University of California, San Diego, La Jolla, CA 92092, USA}

\author{David S.~N. Rupke}
\affiliation{Department of Physics, Rhodes College, Memphis, TN, 38112, USA}
\affiliation{Zentrum für Astronomie der Universität Heidelberg, Astronomisches Rechen-Institut, Mönchhofstr 12-14, D-69120 Heidelberg, Germany}

\author{Wenmeng Ning} 
\affiliation{Department of Astronomy and Astrophysics, University of California, San Diego, La Jolla, CA 92092, USA}
\affiliation{Department of Physics and Astronomy, University of California, Los Angeles, Los Angeles, CA 90095, USA}

\author{Brendan Duong} 
\affiliation{Department of Astronomy and Astrophysics, University of California, San Diego, La Jolla, CA 92092, USA}

\author{Aleksandar M. Diamond-Stanic}
\affiliation{Department of Physics and Astronomy, Bates College, Lewiston, ME, 04240, USA}

\author{Drummond B. Fielding}
\affiliation{Department of Astronomy, Cornell University, Ithaca, NY 14853, USA}
\affiliation{Center for Computational Astrophysics, Flatiron Institute, 162 5th Ave, New York, NY 10010, USA}

\author{James E. Geach}
\affiliation{Centre for Astrophysics Research, University of Hertfordshire, Hatfield, Hertfordshire AL10 9AB, UK}

\author{Ryan C. Hickox}
\affiliation{Department of Physics and Astronomy, Dartmouth College, Hanover, NH 03755, USA}

\author{John Moustakas}
\affiliation{Department of Physics and Astronomy, Siena College, Loudonville, NY 12211, USA}

\author{Gregory H. Rudnick} 
\affiliation{Department of Physics and Astronomy, University of Kansas, Lawrence, KS 66045, USA}

\author{Paul H. Sell}
\affiliation{Department of Astronomy, University of Florida, Gainesville, FL, 32611 USA}

\author{Cameren N. Swiggum} 
\affiliation{University of Wisconsin-Madison, 475 N. Charter St. Madison, WI 53706}
\affiliation{Department of Astrophysics, University of Vienna, Türkenschanzstrasse 17, 1180 Wien, Austria}

\author{Christy A. Tremonti}
\affiliation{University of Wisconsin-Madison, 475 N. Charter St. Madison, WI 53706}

\begin{abstract}

High-velocity outflows are ubiquitous in compact, massive ($\rm M_* \sim$10$^{11} \, M_{\odot}$), $z\sim0.5$ galaxies with extreme star formation surface densities ($\rm \Sigma_{SFR} \sim2000 \, M_{\odot} \ yr^{-1} \ kpc^{-2}$). We have previously detected and characterized these outflows using \mgii\ absorption lines. To probe their full extent, we present Keck/KCWI integral field spectroscopy of the \oii \ and \mgii \ emission nebulae surrounding all of the 12 galaxies in this study. We find that \oii \ is more effective than \mgii \ in tracing low surface brightness, extended emission in these galaxies. The \oii \ nebulae are spatially extended beyond the stars, with radial extent R$_{90}$ between 10 and 40 kpc. The nebulae exhibit non-gravitational motions, indicating galactic outflows with maximum blueshifted velocities ranging from $-$335 to $-$1920 \kmps. The outflow kinematics correlate with the bursty star formation histories of these galaxies. Galaxies with the most recent bursts of star formation (within the last $<$3 Myr) exhibit the highest central velocity dispersions ($\sigma \ga 400$ \kmps), while the oldest bursts have the lowest-velocity outflows. Many galaxies exhibit both high-velocity cores and more extended, slower-moving gas indicative of multiple outflow episodes. The slower, larger outflows occurred earlier and have decelerated as they propagate into the CGM and mix on timescales $\ga$ 50~Myr.
\end{abstract}

\section{Introduction}

Galactic-scale outflows have been increasingly invoked in the last two decades by theoretical models and simulations of galaxy formation to reproduce many of the properties of massive galaxies (e.g., low star formation rates, the relative dearth of low- and high-mass galaxies in the stellar mass function, the black hole–spheroid mass relationship) and the chemical enrichment of the circumgalactic medium (CGM) and intergalactic medium \citep[IGM; e.g.][]{sil98, ker05, tho05, mur10, opp10, dav19, nel19}.
Powered by the energy released from stellar processes and
gas accretion onto supermassive black holes (SMBHs), these outflows provide a mechanism that can
regulate the stellar mass content of galaxies by heating and/or blowing out the gas that fuels star formation and SMBH growth, enriching the large-scale galactic environment with metals. 

Observations have shown that galactic-scale outflows are ubiquitous and span a large range of host galaxy properties \citep[see recent review by][]{vei20}. They are detected in systems at different evolutionary stages, from ultra-luminous infrared galaxies (ULIRGs), through typical main sequence galaxies to quenched elliptical galaxies \citep[e.g.][]{hec00, for03, rup05c, vei13, fio17, bar19, bar20}, and they are observed through different gas phases from high velocity X-ray and UV absorption lines \citep[e.g.][]{blu03, ree03,str07, str09,tom10,mar12, ara13, hec15}, to ionized emission lines \citep[e.g.][]{rub11, har14, rup17, str17, str18, sha22, dut23}, and atomic and molecular emission and absorption  \citep[e.g.][]{rup05, rup05b, fer10, vei13, gea14}. 

While galactic outflows appear to be essential to efficiently quenching star formation, the physical drivers of this ejective feedback remain largely unconstrained. In particular, the relative role of feedback from
stars versus SMBHs in shutting down star formation in massive galaxies is strongly debated \citep[e.g.][]{kor13, vei20}.

Our team uncovered a population of massive (M$_* \sim$10$^{11}$ M$_{\odot}$) galaxies at z = 0.4$-$0.8, originally selected from the Sloan Digital Sky Survey \citep[SDSS;][]{yor00} Data Release 8 \citep[DR8;][]{Aihara11} as young post-starburst galaxies. We refer to these galaxies as the HizEA\footnote{HizEA was initially coined as shorthand for High-$z$ E+A, meaning high redshift post-starburst galaxies. Subsequent work found that most of these galaxies are starbursts (see Section~\ref{subsection:properties}).} sample. Their spectra show strong stellar Balmer absorption from A- and B-stars and weak nebular emission lines, potentially indicating minimal ongoing star formation. These galaxies are driving extremely fast ionized gas outflows as traced by highly blueshifted \mgii \ absorption in $\sim$90\% \citep{dav23} of their optical spectra, with velocities of $\sim 1000-2500$ \kmps, an order of magnitude larger than typical z$\sim$ 1 star-forming galaxies \citep[e.g.,][]{wei09, mar12}. 

Surprisingly, many of these galaxies were detected in the Wide-field Infrared Survey Explorer \citep[WISE;][]{wri10}, and their ultraviolet (UV) to near-IR spectral energy distributions (SEDs) indicate a high level of heavily obscured star formation ($>$50 M$_{\odot}$ yr$^{-1}$; \citealp{dia12}).  Hubble Space Telescope (HST) imaging reveals they are late-stage major mergers with extremely compact central star-forming regions ($R_e \sim$few 100 pc; \citealp{dia12, dia21, sel14}). Combining the star formation rate (SFR) estimates from WISE rest-frame mid-IR luminosities with the small physical sizes from HST imaging leads to exceptionally high SFR surface densities $\Sigma_{SFR}$ $\sim$10$^{3}$ M$_{\odot}$ yr$^{-1}$ kpc$^{-2}$ \citep{dia12}, approaching the theoretical Eddington limit \citep{leh96, meu97, mur05, tho05}. These results reveal that these galaxies are starbursts with dense, dusty star-forming cores, with a substantial fraction of their gas being ejected by powerful outflows \citep{per21,per23}. 
Millimeter observations for two galaxies in the HizEA sample indicate that the reservoir of molecular gas is efficiently consumed by the starburst \citep{gea13}  and ejected in a spatially-extended molecular outflow \citep{gea14}, implying rapid gas depletion times. We find little evidence of ongoing active galactic nucleus (AGN) activity in these systems based on X-ray, IR, radio, and spectral line diagnostics \citep{sel14, per21}. 

The observations are well supported by models of stellar feedback as the primary driver of the observed outflows \citep[e.g.][]{hop12, hop14}.
These compact starburst galaxies have extreme physical conditions and are an ideal laboratory to study the limits of stellar feedback and test whether stellar processes alone can power extreme outflows. While we are catching these galaxies during a brief, dramatic stage in their evolution, most massive galaxies likely undergo such a merger-driven starburst outflow phase, related to star formation quenching. The HizEA sample has a similar space density as ULIRGs and post-starburst galaxies \citep{wha22} and are likely progenitors of massive, compact galaxies at low redshift.

The outflows in these galaxies were discovered through long-slit spectroscopy, which provides limited spatial information and can underestimate the outflow extent, as collimated outflows can be misaligned with the long slit orientation. The most direct insight into large-scale outflows is provided by spatially-resolved spectroscopy [integral-field unit (IFU) observations], which provides the two-dimensional morphology, extent, and velocity field of galactic-scale outflows \citep[e.g.][]{bou07, sha09, new12, bel17, can19, biz19, for19}. In particular, the physical extent of the outflow is key to determining the mass and energy outflow rate and understanding the impact on the host galaxy.

Our team observed one of these compact starburst galaxies (J2118, renamed Makani, ``wind'' in Hawaiian) with the Keck Cosmic Web Imager (KCWI). Makani is a massive (M = 10$^{11.1}$ M$_{\odot}$) star-forming galaxy with $r_e$ = 2.5 kpc \citep{sel14}. The data reveal a spectacular galactic
outflow, $\sim$100 kpc across, traced by \oii \ emission line reaching far into the CGM of the galaxy \citep{rup19}. The \oii \ emission has a bipolar hourglass limb-brightened shape and exhibits a complex kinematic structure with two episodes of ejective feedback that map exactly to two past starburst episodes in the galaxy's star formation history. The KCWI data on Makani directly shows that galactic outflows can feed the CGM, expelling gas far beyond the stars in galaxies. 

In this paper, we present new KCWI observations on an additional 12 of the most well-studied starburst galaxies in the HizEA sample. These IFU data allow us to directly measure morphology, physical extent, and resolved kinematics of the outflows' cool, ionized gas phase. We utilize this data set to probe the structures of the extreme galactic outflows observed in this galaxy sample and investigate the potential impact of these outflows on the evolution of their host galaxies.

The paper is organized as follows: Section~\ref{sec:sample} describes the sample selection and galaxy properties; Section \ref{sec:data} illustrates the observations, data reduction, and our emission line profile fitting method; Section \ref{sec:results} presents our main results; Section~\ref{sec:discussion} discusses the broader implications of our analysis. Our conclusions are summarized in Section \ref{sec:conclusion}. Oscillator strengths and vacuum wavelengths are taken from \cite{mor91, mor03}. Throughout the paper, we assume a standard $\Lambda$CDM cosmology, with H$_0$ = 70 \kmps Mpc$^{-1}$, $\Omega_m$ = 0.3, and $\Omega_{\Lambda}$ = 0.7. All spectra are converted to vacuum wavelengths and corrected for heliocentricity.

\section{Sample}\label{sec:sample}
The parent sample for this work was drawn from the Sloan Digital Sky Survey I (SDSS-I \citealp{yor00}) Data Release 8 (DR8; \citealp{Aihara11}) and includes 121 intermediate redshift (z = $0.4 - 0.8$) starburst galaxies. The sample selection is extensively described in \citet{tre07} and \citet{dav23}. 

We carried out thorough follow-up observations on 50 of these galaxies, prioritizing those with high g-band fluxes and young stellar populations. However, we also included galaxies with older burst ages for comparison, covering a range of mean stellar ages from 4 to 400 Myr. Notably, we did not prioritize galaxies with \ion{Mg}{2} absorption lines. 
We collected ground-based spectroscopy of the 50/121 galaxies with the with the MMT/Blue Channel, Magellan/MagE, Keck/LRIS, Keck/HIRES, Keck/NIRSPEC, Gemini/GMOS, and/or Keck/KCWI \citep{tre07, dia12, sel14, rup19, per21, per23}, X-ray imaging with Chandra for 12/50 targets \citep{sel14}, radio continuum data with the NSF’s Karl G. Jansky Very Large Array (JVLA/VLA) for 20/50 objects \citep{pet20}, millimeter data (ALMA) for 2/50 targets \citep{gea14, gea18}, and optical imaging with HST for 29/50 galaxies \citep[``HST sample''][]{dia12, sel14, dia21}.

For this paper, we selected 12 galaxies from the HizEA sample based on observation scheduling constraints, prioritizing targets that show in their existing long-slit optical spectra \oii \ emission extended either spatially (showing clear physical extension vertically along the slit in the two-dimensional spectra) or spectrally (showing asymmetric broad blue-shifted emission lines). We note that 8 of these galaxies are in the  ``HST'' sample. The galaxies and their basic properties are listed in Table~\ref{table1} and Table~\ref{table2}.

\begin{deluxetable*}{lcCCRRC}
\tablewidth{0pt}
\tablecaption{Galaxy properties\label{table1}}
\tablehead{
\colhead{Object Name}  &  \colhead{\zsys} & \colhead{log(M$_*$/M$_\odot$)} & \colhead{r$_e$} & \colhead{SFR} & \colhead{$\Sigma_{SFR}$} & \colhead{LW Age}\\
\colhead{}  &\colhead{}  & \colhead{} & \colhead{(kpc)} & \colhead{(M$_\odot$ yr$^{-1}$)} & \colhead{(M$_\odot$ yr$^{-1}$} & \colhead{(Myr)} \\
\colhead{}  &\colhead{}  & \colhead{} & \colhead{} & \colhead{} & \colhead{kpc$^{-2}$)} & \colhead{} \\
\colhead{(1)} & \colhead{(2)} & \colhead{(3)} & \colhead{(4)} & \colhead{(5)} & \colhead{(6)} & \colhead{(7)} }
\decimals
\startdata
J0826+4305    & 0.603 & 10.63^{+0.2}_{-0.2}  & 0.173^{+0.075}_{-0.053}  & 184^{+53}_{-41}  &  981     &   22^{+11}_{-5}  \\
J0944+0930    & 0.514 & 10.59^{+0.2}_{-0.2}  & 0.114^{+0.067}_{-0.047}  &  88^{+26}_{-21}  & 1074     &   88^{+48}_{-43}   \\
J1107+0417    & 0.466 & 10.60^{+0.3}_{-0.2}  & 0.273^{+0.194}_{-0.124}  &  72^{+13}_{-14}  & 155      &   10^{+4}_{-2}   \\
J1205+1818    & 0.526 & 10.60^{+0.2}_{-0.3}  & \nodata                  & 147^{+42}_{-34}  & \nodata  &   41^{+14}_{-10}    \\
J1244+4140    & 0.459 & 11.05^{+0.2}_{-0.1}  & \nodata                  &  82^{+26}_{-18}  & \nodata  &  156^{+60}_{-56}   \\
J1341$-$0321  & 0.661 & 10.53^{+0.2}_{-0.1}  & 0.117^{+0.040}_{-0.032}  & 151^{+34}_{-23}  & 1755     &   14^{+6}_{-3}   \\
J1500+1739    & 0.576 & 10.88^{+0.2}_{-0.2}  & \nodata                  & 157^{+26}_{-26}  & \nodata  &   24^{+8}_{-6}   \\
J1506+5402    & 0.436 & 10.60^{+0.2}_{-0.2}  & 0.168^{+0.076}_{-0.054}  & 116^{+32}_{-25}  &  652     &   13^{+6}_{-2}  \\
J1558+3957    & 0.402 & 10.42^{+0.3}_{-0.3}  & 0.778^{+0.383}_{-0.244}  &  84^{+16}_{-15}  &   22     &   44^{+14}_{-10}   \\
J1613+2834    & 0.449 & 11.12^{+0.2}_{-0.2}  & 0.949^{+0.274}_{-0.207}  & 172^{+36}_{-36}  &   30     &   72^{+33}_{-26}   \\
J1622+3145    & 0.441 & 10.62^{+0.2}_{-0.2}  & \nodata                  & 151^{+51}_{-53}  & \nodata  &   34^{+17}_{-8}   \\
J1713+2817    & 0.576 & 10.89^{+0.1}_{-0.1}  & 0.173^{+0.030}_{-0.030}  & 229^{+53}_{-98}  & 1218     &  134^{+34}_{-24} \\
\hline
J2118+0017\tablenotemark{a} & 0.459 & 10.95^{+0.1}_{-0.1} & 2.240^{+0.400}_{-0.400} & 230^{+92}_{-76} & 7  &    95^{+37}_{-27} 
\enddata
\tablecomments{ Col 2: Galaxy systemic redshift; Col 3: Stellar mass from Prospector; Col 4: Effective radii from HST; Col 5: SFRs from Prospector; Col 6: SFR surface densities estimated using columns (4) and (5); Col 7: Light-weighted ages of the stellar populations younger than 1 Gyr.}
\tablenotetext{a}{J2118+0017, or Makani, is not part of our sample but we include it as a comparison.}
\end{deluxetable*}

\subsection{Galaxy properties}\label{subsection:properties}

A suite of relevant galaxy properties derived for our sample is listed in Table~\ref{table1}. 
To determine accurate outflow velocities, we derived precise galaxy systemic redshift (\zsys) from the stellar continuum fits as described in \citet{dav23}. 
The effective radii (r$_e$) estimates for the eight galaxies in this work that are part of the ``HST sample'' are extensively discussed in \citet{dia12, dia21}. 

We use the Bayesian SED code Prospector \citep{leja19, johnson21} to fit the combined broad-band UV -- mid-IR photometry and optical spectra to derive stellar masses (M$_*$) and star formation rates (SFR). For a complete description of this procedure see \citet{dav23}. In brief, we include the 3500 - 4200 \AA\ spectral region in the SED fit as it covers many age-sensitive features (e.g., D4000, H$\delta$). 
We utilize the Flexible Stellar Populations Synthesis code \citep[FSPS;][]{con09}, to generate simple stellar population (SSP) models adopting a Kroupa IMF \citep{kro01} and the MIST isochrones \citep{cho16} and the C3K stellar theoretical libraries \citep{cho16,byr23}.
We obtain the best-fit parameters and their errors from the 16th, 50th, and 84th percentiles of the marginalized probability distribution function. The combined photometry and spectra are well fit by these models (see \citealp{dav23} for examples of the SED fitting). However, the low signal-to-noise ratio (SNR) of the WISE W3 and W4 photometry and the limited infrared coverage of the SED yield poorly constrained dust emission properties of our sample. This results in relatively tight constraints on the M$_*$ ($\pm$0.15 dex) and slightly larger errors on the SFR ($\pm$0.2 dex). M$_*$ represents the present-day stellar mass of the galaxy (after accounting for stellar evolution) and not the integral of the star formation history (i.e. total mass formed). We report in Table~\ref{table1} the SFRs estimated from the star formation history (SFHs; see Fig.~\ref{fig:sfh}), averaged over the final 100 Myr. This is the characteristic timescale for which UV and IR star formation indicators are sensitive \citep{ken12}. We compute the light-weighted age of the stellar populations younger than 1 Gyr, using the light contribution at 5500 \AA, as we are interested in the most recent SFH. These $<$1 Gyr light-weighted ages more closely reproduce the timescale of the peak SFR than the mass-weighted ages. We utilize light-weighted ages rather than the age since the burst as the light-weighted ages are more robust to modifications in our modeling process.  However, there may be systematic errors associated with the stellar population models we assume.  For example, uncertainties in the treatment of  Wolf-Rayet stars and high mass binary evolution can largely affect the UV spectra of galaxies with young stellar populations \citep[e.g.][]{eld16}. The detailed analysis needed to make a quantitative estimation of the systematic errors on the light-weighted ages is beyond the scope of this work. The light-weighted ages for our sample are listed in Table~\ref{table1}.

\subsection{Makani}\label{makani}
Our team observed the galaxy Makani with KCWI and uncovered a $\sim$100 kpc nebula surrounding the galaxy \citep{rup19}. The KCWI observation revealed two distinct galactic wind episodes traced by \oii \ emission. Combining the analysis of its morphology, kinematics, and stellar populations, we inferred that Episode I was powered by a star formation episode 400 Myr in the past, and includes most of the outer 20$-$50 kpc of the wind. This wind has slow projected speeds $\sim$100 \kmps and linewidths $\sigma$ = 200 \kmps, with a bipolar limb-brightened shape characteristic of bipolar outflows. Episode II was powered by star formation 7 Myr ago and consists of a fast wind with maximum speeds exceeding 2000 \kmps. Most of the Episode II wind is within 20 kpc of the host galaxy, though there is a faint southern extension to 40 kpc. 

\begin{deluxetable*}{ccrcrc}
\tabletypesize{\small}
\tablecaption{Sample Observations\label{table2}}
\tablehead{
\colhead{Object Name } & \colhead{RA} & \colhead{Dec} & 
\colhead{Obs. Date} &\colhead{Exp. time} & \colhead{$\lambda$ coverage}  \\
\colhead{}& \colhead{J2000} & \colhead{J2000} & \colhead{}& \colhead{(min)}  & \colhead{(\AA)}  \\
\colhead{(1)} & \colhead{(2)} & \colhead{(3)} & \colhead{(4)} & \colhead{(5)} & \colhead{(6)}  }
\startdata
J0826+4305   &126.66006 & 43.091498 &  Dec 29 2019  &  60               & 4170 - 6330 \\
J0944+0930   &146.07437 &  9.505385 &  Dec 29 2019 - Mar 18 2021 & 140  & 3770 - 5930  \\
J1107+0417   &166.76196 &  4.284098 &  Dec 29 2019 - Mar 18 2021 & 100  & 3770 - 5930 \\
J1205+1818   &181.46292 & 18.313869 &  May 24 2020 - Mar 18 2021 & 120  & 3670 - 5830 \\
J1244+4140   &191.09048 & 41.674771 &  Mar 18 2021 - May 07 2021 & 100  & 3670 - 5830 \\
J1341$-$0321 &205.40333 & -3.357019 &  Dec 29 2019 - Mar 18 2021 &  80  & 4170 - 6330 \\
J1500+1739   &225.17823 & 17.655084 &  May 24 2020               & 100  & 4070 - 6230 \\
J1506+5402   &226.65124 & 54.039095 &  May 24 2020 - May 07 2021 & 100  & 4070 - 6230 \\
J1558+3957   &239.54683 & 39.955787 &  Mar 18 2021               &  24  & 3670 - 5830 \\
J1613+2834   &243.38552 & 28.570772 &  May 24 2020 - May 07 2021 & 140  & 3670 - 5830 \\
J1622+3145   &245.69627 & 31.759129 &  May 07 2021 - May 24 2020 & 100  & 3670 - 5830 \\
J1713+2817   &258.25160 & 28.285630 &  May 07 2021               &  40  & 4070 - 6230 \\  
\enddata
\tablecomments{ Col 4: observation date; Col 5: exposure time; Col 6: observed wavelength coverage.}
\end{deluxetable*}

\section{OBSERVATIONS AND DATA PROCESSING}\label{sec:data}
This section briefly illustrates our data and describes the method and assumptions adopted for the line profile. The line fitting results for each galaxy in kinematics map that we present in Figures~\ref{fig:quad_1}$-$\ref{fig:1613_kin}. Examples of the line fits for each galaxy in our sample are visible in Figures~\ref{fig:inset1},~\ref{fig:inset2}, and \ref{fig:inset_3}.

\subsection{KCWI observations}

We obtained rest-frame near UV-optical spectra of 12 starburst galaxies spanning the emission redshifts 0.4 $< z <$ 0.7, using the Keck Cosmic Web Imager \citep[KCWI;][]{mor18} on the Keck II telescope over the nights of 29 December 2019, 24 May 2020, 12 February 2021, 18 March 2021, and 7 May 2021. Table~\ref{table2} provides some basic information about our new observations. KCWI was configured using the blue low-dispersion (BL) grating and medium slicer, which provides a spectral resolution of $R = 1800$, a spaxel size of $0.29'' \times 0.69''$, and a field of view (FOV) of $20'' \times 16''$ per pointing. 
We used a central wavelength ($\lambda_{c}$) of 4800, 4900, 5200, and 5300 \AA \ and a detector binning of $2 \times 2$, yielding a rest-frame wavelength coverage from $\lambda_{rest} \sim 2400 -2850$ to $3800-4350$~\AA\, depending on redshift. The seeing was quite consistent through the nights $\sim 1''$, corresponding to a projected distance of $5 - 7$ kpc at the redshifts of our targets. The angular size of our targets fits KCWI FOV. Each galaxy was observed for 24 to 140 minutes, consisting of individual exposures of 12 or 20 minutes. Individual exposures were dithered $0.35 ''$ along slices to subsample the output spaxels.   
We adopted a position angle close to the parallactic angle at the moment of the observation. At the end of the data reduction, we rotated each data cube to a position angle of zero using a custom Python routine.

\subsection{Data Reduction}
We reduced the data of the individual science exposures with the standard KCWI Data Extraction and Reduction Pipeline (KDERP\footnote{https://github.com/Keck-DataReductionPipelines/ KcwiDRP}) written in the Interactive Data Language (IDL). We followed all pipeline stages. Sky subtraction was performed using a manual selection of a sky mask region in each frame. For each exposure, we utilized a standard star observed with the same setup for the flux calibration.
The final outputs are data cubes of two spatial dimensions and one spectral dimension. We used the IDL library IFSRED \citep{rup14} to resample the data onto $0.29'' \times 0.29''$ spaxel grids using the routine IFSR\_KCWIRESAMPLE; align the exposures for each target by fitting the galaxy centroid (IFSR\_PEAK); and generate a mosaic of the data (IFSR\_MOSAIC). 
The resulting stacked and resampled data cubes have dimensions of 63 $\times$ 72 spaxels, covering $18'' \times  21''$ and corresponding to a projected physical size of $116 \times 132$ kpc at $z = 0.5$. 

\subsection{Emission-line fitting}\label{subsec:emlfit}

We model the spectrum in each spaxel of the data cubes using the IDL library IFSFIT \citep{rup14a}. This library employs Penalized Pixel-Fitting \citep[PPXF;][]{cap12} to fit the stellar continuum, and MPFIT \citep{mar09} to fit a user-defined number of Gaussian profiles to the emission lines.

IFSFIT masks spectral regions containing emission lines before fitting the continuum. Rather than independently fitting multiple stellar models to each spaxel, we create a single starlight template from a fit to a spatially-integrated spectrum. We sum light from the inner regions of each galaxy. IFSFIT produces a fit to the spatially-integrated spectrum using the high-resolution stellar population synthesis (SPS) model \citet{gon05} assuming solar metallicity and Legendre polynomials that account for residuals from imperfect calibration (e.g. scattered light, sky subtraction). We utilize the resulting template to model and subtract the continuum in each spaxel.

Our KCWI datacubes cover the \oii$\lambda\lambda$3726, 3729 and \mgii$\lambda\lambda$2796, 2803 doublets. However, we fit only the \oii \ emission lines as the \mgii \ signal in individual spaxels is too weak to perform robust Gaussian fits. We come back to the \mgii \ in Section~\ref{sec:eml}.
Since the \oii$\lambda\lambda$3726, 3729~\AA \ doublet is not spectrally resolved, the flux ratio of \oii \ 3726/3729 is fixed to 1.2.
We model the \oii \ doublet in each spaxel with one or two Gaussian functions, depending on the complexity of the emission profiles and the SNR. Model line profiles are convolved with the spectral resolution before fitting. We first perform the fit allowing only one Gaussian component for the \oii \ emission lines. Then, we visually explore the resulting best fits and we re-fit the spectra employing two Gaussian components when the improvement in $\chi^2$ is statistically significant, accounting for the additional free parameters. 
To validate our results, we apply the F-test to compare the fits by analyzing the ratio of their variances. We ensure that the inclusion of the second Gaussian component results in a statistically significant improvement at the 3$\sigma$ level (p $<$ 0.00135). This criterion ensures that a substantial improvement in the model fit justifies the additional complexity.
After fitting with the final number of components, \oii \ emission lines with a significance of less than $3\sigma$ in the total flux are set to zero. We find the \oii \ lines require two Gaussian components only in the central regions of J1613 and J1622.

Three of the galaxies have slight additions to the fitting procedure: 1) For J0826, IFSFIT fails to fit 57 spaxels, representing around 10 percent of spaxels with valuable flux and \oii \ SNR $>$ 3. We set the kinematics values for these spaxels to the median values of the eight surrounding spaxels with good fits. 2) J1613 and J1622 are the two sources in our sample that require a second Gaussian component to properly fit the \oii \ emission in the central region of the galaxy. We select the ``first'' and ``second'' components according to their velocity dispersion, with the first component being the narrowest.

\subsection{Emission-line analysis}\label{em_an}
\begin{figure*}[htp!]
 \centering
 \includegraphics[width=0.9\textwidth]{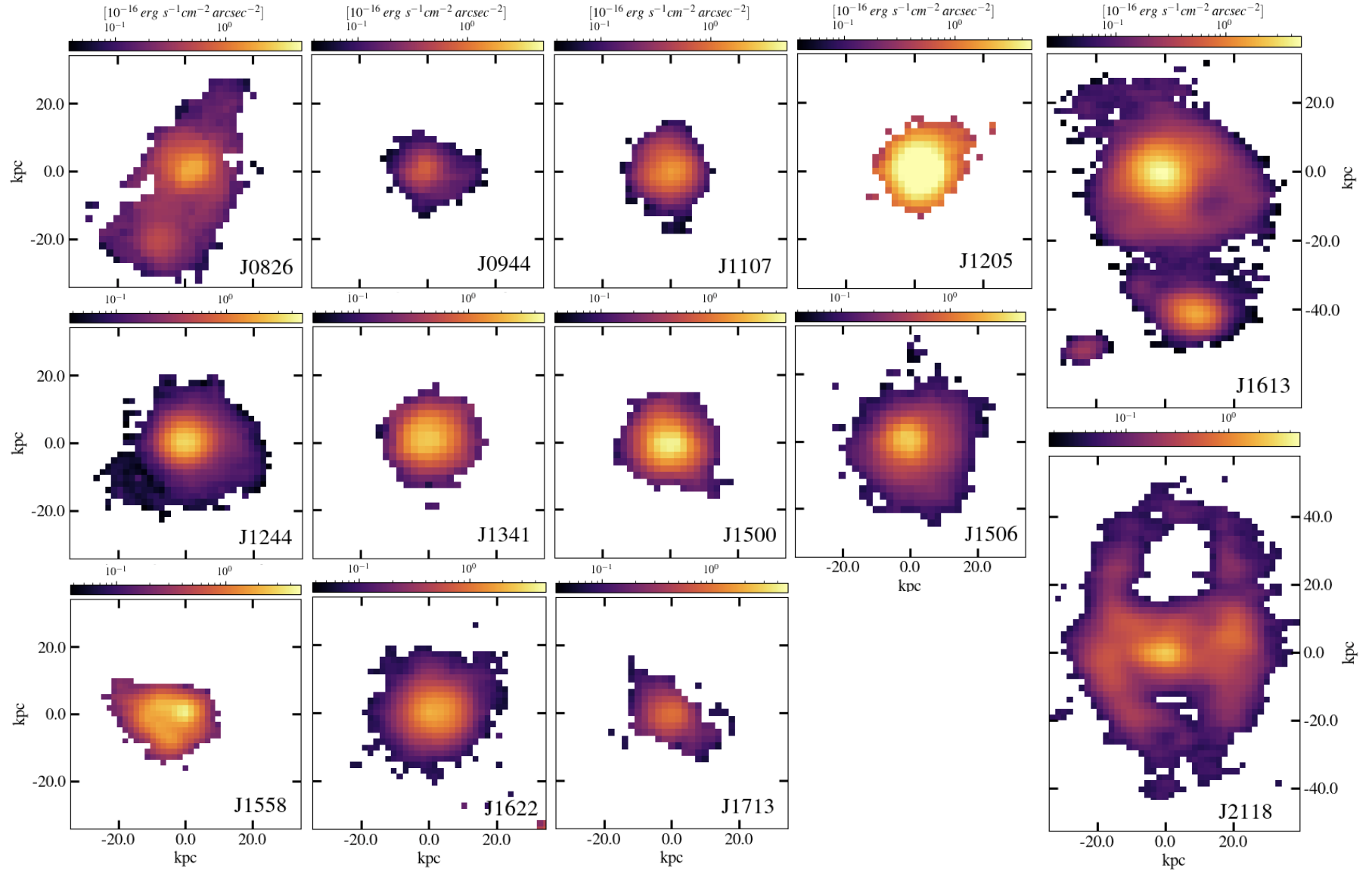}
 \caption{Colors show observed-frame \oii \ surface brightness for the galaxies in our sample and the galaxy Makani (J2118) as a comparison. On each panel, we show the area with a signal detected above a signal-to-noise threshold of four. The axes are labeled in kpc from the brightest spaxel. Note we use the same logarithm color scale for the whole sample to ease comparison.} \label{fig:o2_sb}
\end{figure*}

We use the \oii$\lambda\lambda$3726,3729 model line profiles derived in Section~\ref{subsec:emlfit} to create emission line maps to probe the spatial extent, morphology, and spatially-resolved kinematics of the ionized gas for each galaxy in our sample. We obtain:

\begin{itemize}
    \item Surface brightness (SB) maps: the \oii \ SB maps (Figure~\ref{fig:o2_sb}) are obtained by dividing the modeled continuum-subtracted \oii \ flux in each spaxel by the spaxel area. We only consider spaxels with a signal-to-noise ratio above four. 
    Spaxels are squares with 0.2914\arcsec per side.

    \item Surface brightness radial profiles: the \oii \ (Figure~\ref{fig:sb_profile}) SB radial profiles are obtained azimuthally-averaging the SB over spaxels in bins of radial distance from the brightest spaxel. 
    Errors are the standard deviation from the mean.  

    \item Dust extinction: we correct the emission line fluxes for galaxy intrinsic dust extinction by comparing the Balmer decrement (H$_{\alpha}$/H$_{\beta}$ or H$_{\gamma}$/H$_{\beta}$) with the expected CaseB value \citep[2.86 or 0.47; ][]{ost89}. Galaxies with Balmer decrement less than the expected value (but consistent with it within the uncertainties) are assumed to have zero extinction. 
    We also adopt the Galactic extinction curve from \citet{car89}. We use ancillary data from \citet{per21} for 8 galaxies in our sample (J0826, J0944, J1107, J1341, J1506, J1613, J1622, and J1713) and SDSS data for the remaining 4 (J1205, J1244, J1500, and J1558).
    
    \item \oii \ luminosity: we create a spatially-integrated spectrum (Figure~\ref{fig:integrated}) by summing the continuum-subtracted flux in a rectangular aperture containing the 4$\sigma$ \oii \ nebula. We fit Gaussian profiles to the spatially-integrated \oii \ doublet emission lines and correct the modeled line flux for Galactic and galaxy intrinsic dust extinction. Finally, we convert the corrected flux into luminosity using the luminosity distance at the galaxy's redshift. The \oii \ luminosity values are reported in Table~\ref{table3}.
    
\end{itemize}

\subsection{MgII emission lines}\label{em_mg_sb}

 Our KCWI datacubes also cover the \mgii$\lambda\lambda$2796, 2803 \ doublet. Unlike the strong \oii \ emission, the \mgii \ signal in individual spaxels is too weak to perform robust Gaussian fits to the emission lines. Therefore to create emission line maps to study the \mgii \ spatial distribution we proceed as follows. For each galaxy we use the stellar continuum-subtracted datacube to produce a spatially-integrated spectrum spanning the full nebula and one spanning the central 5 $\times$ 5 spaxels (see Appendix~\ref{app:A}). We utilize both integrated spectra to identify the spectral wavelength range that includes the \mgii \ emission profile and avoids \mgii \ absorption.
We then integrate the datacube in each spaxel over the wavelength interval determined (marked in Figure~\ref{fig:integrated}) to obtain the \mgii \ emission line flux in that spaxel. Finally, we visually inspect the spectra in individual spaxels to help identify the spatial regions in the flux map where the \mgii \ emission looks real, such that the signal is persistent in at least a few adjacent spaxels. 

For each spaxel we sum in quadrature the pixels in the co-added variance cube over the same spectral window and calculate the  \mgii \ emission detection significance using

\begin{equation}
    S_{j,k} = \frac{\sum_{i} f_{i,j,k}}{\sqrt{\sum_{i} \sigma_{i,j,k}^2}}
\end{equation}
where $S_{j,k}$ is the significance of the \mgii \ emission in spaxel (j,k), $f_{i,j,k}$ and $\sigma_{i,j,k}^2$ are the flux and variance, respectively, of the $i$th spectral voxel in the (j,k) spaxel.

We obtain:

\begin{itemize}

\item The \mgii \ SB maps (Figure~\ref{fig:mg2_sb}) are obtained by dividing the \mgii \ flux in each spaxel by the spaxel area. We only consider spaxels with a signal-to-noise ratio above one. 
    Spaxels are squares with 0.2914\arcsec per side.

\item From the wavelength integrated spectra we measure the \mgii \ rest-frame equivalent width and \mgii \ luminosity, reported in Table~\ref{table4}. 

\item Surface brightness radial profiles are obtained by azimuthally-averaging the SB over spaxels with a S/N threshold of 1$\sigma$ in bins of radial distance from the brightest spaxel. 

\end{itemize}

\begin{figure*}[htp!]
 \centering
 \includegraphics[width=0.9\textwidth]{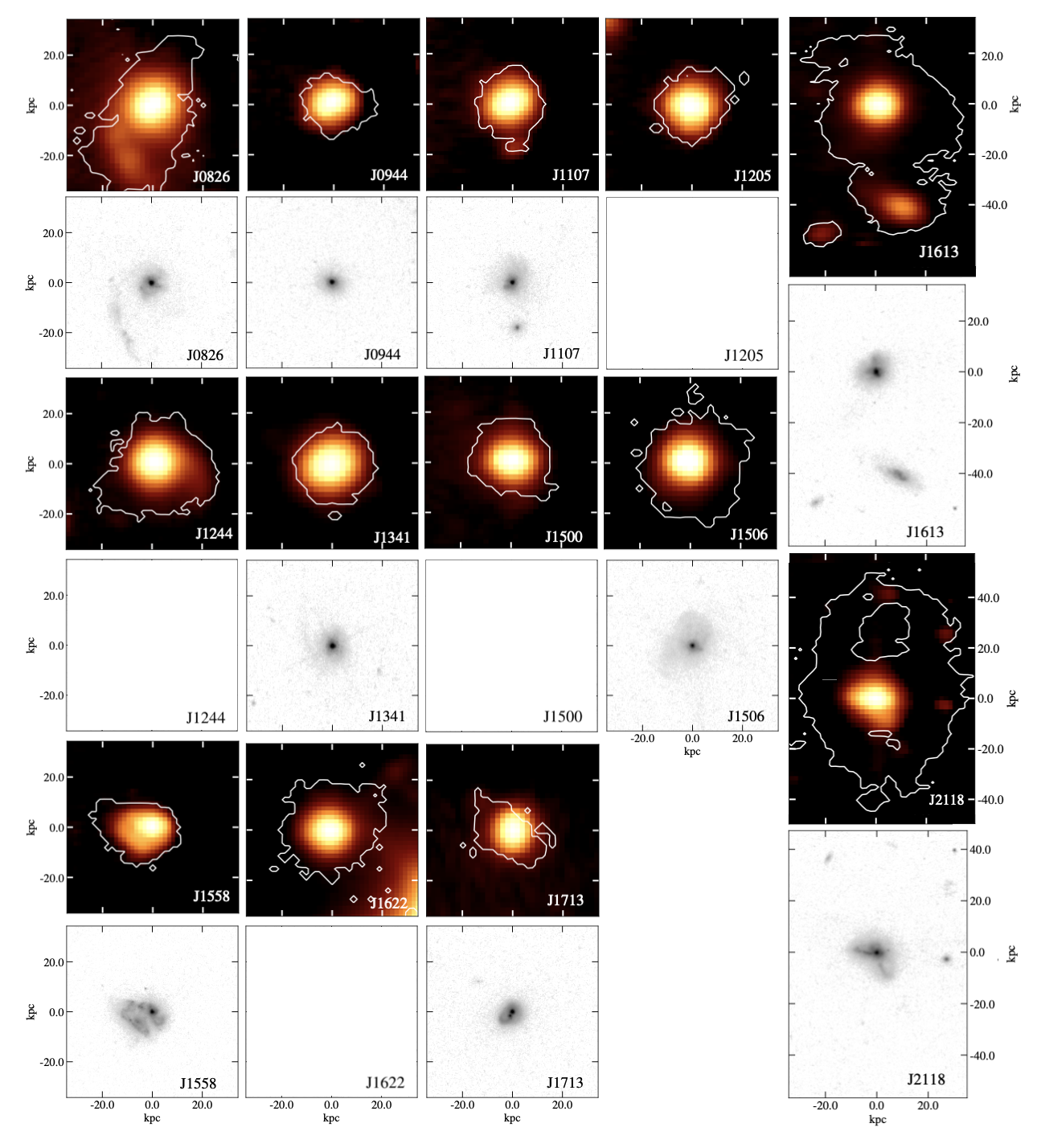}
 \caption{The panels with a black background show a continuum image illustrating stellar emission at observed-frame $4000-5500$ \AA \ for the galaxies in our sample and the galaxy Makani (J2118). On each panel, the white contour represents the 4$\sigma$ \oii \ emission as presented in Figure~\ref{fig:o2_sb}. We use a logarithm color scale. The panels with a white background show HST/WFC3 F814W (restframe V-band at these redshifts) observations of eight galaxies in our sample and the galaxy Makani (J2118). The axes are labeled in kpc from the brightest spaxel. North is up and east is left.
 } \label{fig:fig2}
\end{figure*}

\section{Results}\label{sec:results}

\subsection{Morphology}\label{sec:mor}
KCWI is exceptional at mapping low surface brightness extended emission, which allows us to trace the morphology of the ionized gas nebulae in our sources to large distances. Figure~\ref{fig:o2_sb} shows the  \oii$\lambda\lambda$3726,3729 surface brightness maps for the galaxies in our sample. In each panel we apply a signal-to-noise threshold per spaxel of S/N = 4 and the origin is centered on the brightest spaxel.
Throughout this paper we refer to the angular separation between extreme points in a contiguous detection region as the maximum radial extent.

The \oii \ emission in these galaxies reveals nebulae spanning a wide range of shapes and physical extents. Within the general context of \oii \ emitters, the sizes of these nebulae are remarkable, with areas ranging from 500 to 3300 kpc$^2$ (above an average surface brightness limit of 2$\times$10$^{-18}$ erg s$^{-1}$ cm$^{-2}$ arcsec$^{-2}$) and maximum radial extents between 10 and 40 kpc. These are among the largest \oii \ nebulae detected around isolated field galaxies \citep{bri15, yum17}. Moreover, their median luminosity of 3$\times 10^{42}$ erg s$^{-1}$ is a factor two higher than the break in the luminosity function, $L^*$, at the median redshift of our sample $z\sim0.5$ \citep{zhu09}.

Figure~\ref{fig:fig2} (black background panels) presents stellar continuum images for the galaxies in our sample, created by integrating the spectrum at each spaxel in a $\sim 1000$ \AA \ emission line-free region. The observed-frame wavelength range for this integration is between $4000$ and $5500$ \AA, depending on the galaxy's redshift. All galaxies show a bright, compact central stellar source surrounded by diffuse stellar emission. The light distribution often exhibits clear tidal tails or disturbed morphologies, indicative of major or minor mergers. 

Figure~\ref{fig:fig2} (white background panels) displays cutouts from HST/WFC3 F814W (restframe V-band at these redshifts) observations for 9 of the galaxies in our sample from \citet{sel14}. The authors explore the nature of the extended diffuse light using quantitative image analysis performed with GALFIT and find the presence of tidal debris in all of these galaxies except J0944. However, given the shallow depth of the HST images, they cannot rule out the presence of such features in J0944. Among the features immediately evident when comparing the KCWI stellar continuum and the HST images are the prominent tidal tail extending to the southeast of J0826, the tidal tail extending to the northeast of J1341, the asymmetric diffuse light indication of an ongoing merger in J1558, and the double core in J1713 (as well as the tidal features previously highlighted in J2118).

To facilitate comparison between the extent of the stellar and gas emission, in each panel (with black background) in Figure~\ref{fig:fig2}, we display a white contour representing the 4$\sigma$ \oii \ emission, as illustrated in Figure~\ref{fig:o2_sb}. The brightest \oii \ emission coincides with the peak of the stellar continuum. The \oii \ and stellar continuum emission overlap spatially, but in most galaxies the \oii \ is far more extended.

We identify four morphology categories in which to group the galaxies in our sample that have common characteristics. Five galaxies (J0944, J1107, J1205, J1341, and J1500) show fairly round \oii \ emission with an average maximum radial extent of $\sim 15$ kpc. These galaxies have the least extended \oii \ nebulae in our sample, and the \oii \ emission has high spatial overlap with the stellar continuum emission, though in J0944, J1205, and J1500  the ionized gas extends a few kpc beyond the stars. Three galaxies (J1244, J1506, and J1622) exhibit somewhat rounded and larger \oii \ nebulae, with an average maximum radial extent of $\sim 23$ kpc. The \oii \ emission in these galaxies is not perfectly centered on the stellar continuum and extends from 5 to 15 kpc beyond it.  Two galaxies (J0826 and J1613) host the largest \oii \ nebulae in our sample (other than J2118), with a maximum radial extent of  $\sim$40 kpc. Their \oii \ emission extends considerably farther than the stellar continuum and merger debris. The remaining two galaxies (J1558 and J1713) are peculiar, as both show strongly asymmetric and highly elongated \oii \ nebulae, almost twice as long in one dimension as in the other. J1558 is an ongoing merger, and the \oii \ emission extends to the northeast of the galaxy, following the merger debris and $\sim$ 9 kpc beyond the tidal features. J1713 is a type II AGN candidate \citep{sel14, per21}, and the \oii \ emission extends $\sim$ 6 kpc to the northeast and $\sim$ 4 kpc to the southwest beyond the stars in the galaxy. While this \oii \ nebula may be elongated along the direction of AGN jets or an AGN-driven outflow, we do not have information about the presence of jets in this galaxy.

Interestingly, for at least four galaxies in our sample (J0826, J1244, J1506, and J1613) the \oii \ nebula both overlaps with tidal tails and debris due to recent merger events, while also extending in the {\it opposite} direction, beyond the stellar continuum 
$\sim$10 to 15 kpc, depending on the galaxy.

The largest \oii \ nebula in our sample belongs to J1613 and is similar to the one observed in Makani (J2118) in several respects. The outflows in both galaxies have similar physical scales and limb-brightened bubbles surrounding evacuated regions. In particular, we observe hollow regions in the southwest, southeast, and northeast of J1613, within a radius of 20 kpc from the compact starburst. An important way in which J1613 {\it differs} from Makani is its environment; it is not isolated (which may be why it does not have as symmetric of a nebula) and is interacting with a smaller galaxy $\sim$ 40 kpc away \citep{sel14}, at the same redshift.

\begin{figure*}[htp!]
 \centering
 \includegraphics[width=0.9\textwidth]{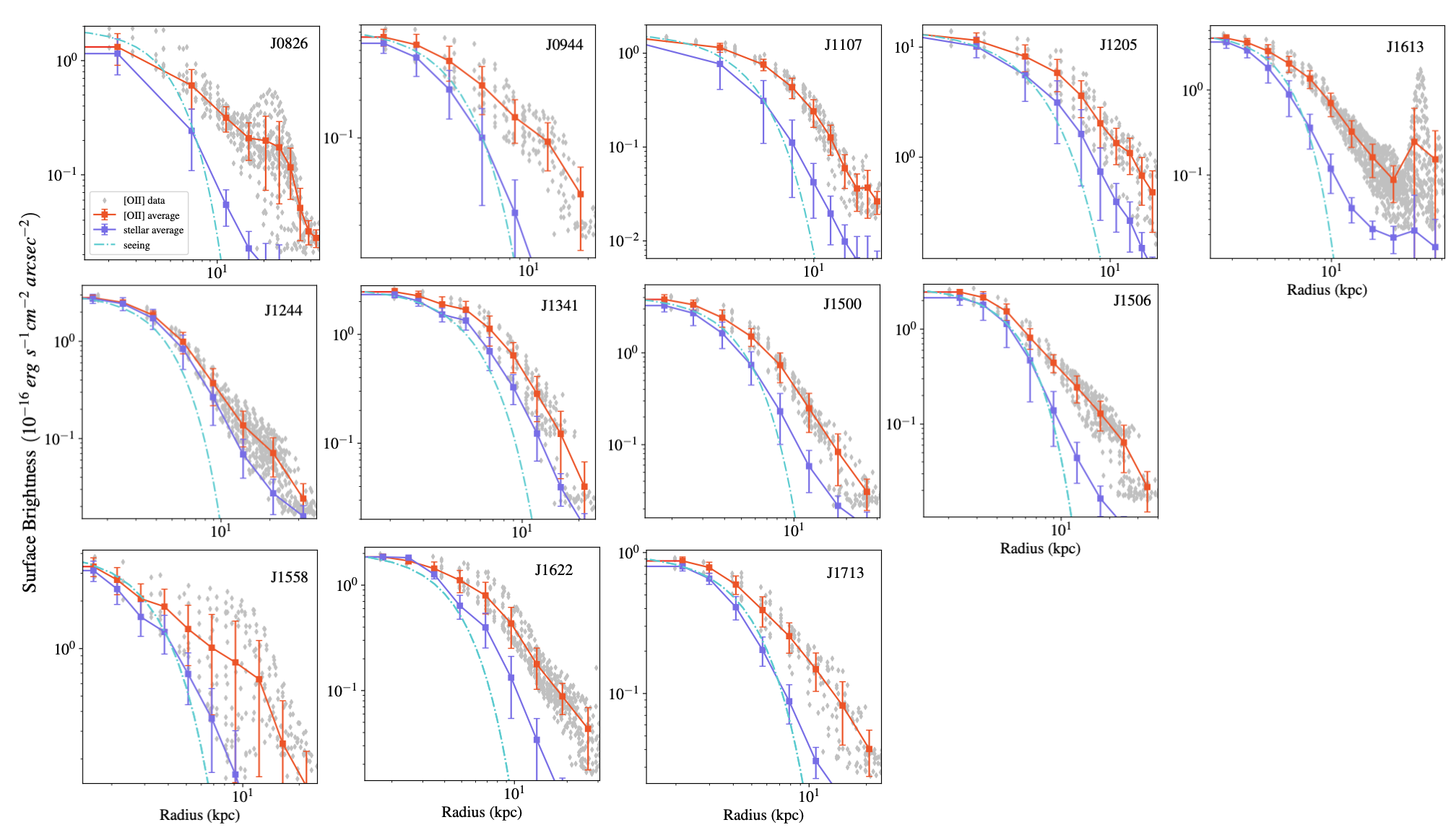}
 \caption{Radial surface brightness profile of the \oii \ emission. For each galaxy, the zero point is set to be the brightest spaxel in the surface brightness map, i.e. the origin of each panel in Figure~\ref{fig:o2_sb}. The red line shows the azimuthally-averaged \oii \ radial surface brightness profile, and the grey diamonds are the \oii \ surface brightness data at every spaxel. The purple line shows the radial profile of the stellar continuum, normalized to the peak of the \oii \ profile, and the cyan line shows the extent of the ground-based seeing, also normalized to the \oii. Errors are the standard deviation from the mean.}
 \label{fig:sb_profile}
\end{figure*}

\begin{deluxetable*}{crCCDDDCCCC}
\tablecaption{[OII] Outflow Properties\label{table3}}
\tablehead{
\colhead{Object Name}  & \colhead{Area}  & \colhead{L\oii} & \colhead{L\oii$_\mathrm{dust}$} & \twocolhead{EW} & \twocolhead{R$_{50}$} & \twocolhead{R$_{90}$}  & \colhead{max($v_{02}$)} & \colhead{min($v_{98}$)} & \colhead{med($\sigma_c$)}\\
\colhead{} & \colhead{(kpc$^2$)} & \colhead{(erg s$^{-1}$)} & \colhead{(erg s$^{-1}$)} & \twocolhead{(\AA)} & \twocolhead{(kpc)} & \twocolhead{(kpc)} & \colhead{(\kmps)} & \colhead{(\kmps)} & \colhead{(\kmps)}\\
\colhead{(1)} & \colhead{(2)} & \colhead{(3)} & \colhead{(4)} & \twocolhead{(5)} & \twocolhead{(6)} & \twocolhead{(7)} & \colhead{(8)} & \colhead{(9)} & \colhead{(10)}
}
\decimals
\startdata
J0826+4305    & 1565 & 2.6 \times 10^{42} & \nodata            & 13.75 & 10.68 & 25.82 & 917  & -693   & 194 \\
J0944+0930    & 501  & 3.5 \times 10^{41} & 1.4 \times 10^{42} &  4.06 &  5.66 & 12.51 & 753  & -718   & 218 \\
J1107+0417    & 599  & 9.9 \times 10^{41} & 2.5 \times 10^{42} &  8.98 &  5.38 & 11.20 & 818  & -712   & 299 \\
J1205+1818    & 609  & 6.2 \times 10^{42} & 5.3 \times 10^{43} & 11.03 &  5.42 & 11.01 & 716  & -500   & 156 \\
J1244+4140    & 1417 & 1.2 \times 10^{42} & \nodata            & 12.12 &  6.13 & 17.54 & 495  & -330   &  74 \\
J1341$-$0321  & 718  & 2.3 \times 10^{42} & 5.6 \times 10^{42} &  8.43 &  6.38 & 12.24 & 1273 & -1822  & 509 \\
J1500+1739    & 869  & 2.2 \times 10^{42} & 9.0 \times 10^{42} & 16.99 &  5.66 & 12.01 & 1328 & -890   & 375 \\
J1506+5402    & 1437 & 2.2 \times 10^{42} & 3.5 \times 10^{42} & 10.68 &  7.90 & 17.41 & 829  & -880   & 237 \\
J1558+3957    & 607  & 1.9 \times 10^{42} & 2.5 \times 10^{42} & 15.43 &  8.00 & 15.00 & 556  & -640   & 127 \\
J1613+2834    & 3322 & 2.7 \times 10^{42} & 6.3 \times 10^{42} & 30.84 & 10.43 & 41.63 & 1091 & -1072  & 414 \\
J1622+3145    & 1276 & 1.2 \times 10^{42} & 2.4 \times 10^{42} & 10.60 &  7.53 & 16.19 & 786  & -823   & 212 \\
J1713+2817    & 713  & 5.1 \times 10^{41} & 1.8 \times 10^{42} &  4.59 &  6.00 & 13.96 & 800  & -389   & 129 \\
\hline
J2118+0017    & 5683 &2.8 \times 10^{42}  & 1.9 \times 10^{43}\tablenotemark{a} & 39.76  & 18.00    & 39.00    & 1788 & -2518 & 408
\enddata 
\tablecomments{Col 2: Area of the 4$\sigma$ \oii \ nebula; Col 3: \oii \ luminosity corrected for Galactic extinction; Col 4: \oii \ luminosity corrected for Galactic and galaxy intrinsic dust extinction; Col 5: rest-frame \oii \ equivalent width; Col 6: radius containing 50\% of the light in the surface brightness radial profile; Col 7: radius containing 90\% of the light in the surface brightness radial profile; Col 8: maximum redshifted $v_{02}$; Col 9: minimum blueshifted $v_{98}$; Col 10: median $\sigma$ calculated in the central 5$\times$5 sapxels. The median and maximum values are calculated after clipping the velocity distributions at 3$\sigma$ to eliminate outliers.}
\tablenotetext{a}{We use an extinction model derived from \citet{rup23}: E($B-V$)$=0.5$ for $r < 25$~kpc and 0 otherwise, and $R_\mathrm{V}=3.1$.}
\end{deluxetable*}

\subsection{Radial Profiles}\label{sec:rad_profile}
We next investigate the radial surface brightness (SB) profiles of the \oii \ nebulae to quantify their extent and compare it with that of the stellar continuum. Despite the clear asymmetries in some of the nebulae, we use the standard approach in the literature of measuring azimuthally-averaged SB profiles, which facilitates comparison with previous work. We do not apply a S/N cut to the SB maps for this calculation.

The resulting azimuthally-averaged \oii \ radial SB profiles for each nebula are presented as red lines in Figure~\ref{fig:sb_profile}, while the grey diamonds are the \oii \ SB data in every spaxel. For comparison in each panel we display the average radial profile of the stellar continuum (purple line) and the seeing (cyan line) profile, both of which are normalized to the peak of the \oii \ profile.

In all galaxies the \oii \ emission is more extended than the stars and considerably more extended than the seeing.
Despite the different morphologies, sizes, and luminosities, most of the \oii \ SB profiles look similar to each other, relatively shallow in the central regions with a decrease at increasing radial distances that is less steep than the stellar continuum profile. Several \oii \ SB profiles contain a large amount of scatter at a given radial distance, which is indicative of the asymmetry of the \oii \ nebula. In particular,  J0826, J1558, and J1613 have a high amount of scatter.

To characterize the \oii \ emission we calculate the radii containing 50\% and 90\% of the light in the \oii \ SB radial profile, R$_{50}$ and R$_{90}$, integrating from the center of the nebula outward.
These values are reported in Table~\ref{table3}. We deconvolve measurements of R$_{50}$ with the seeing, while seeing corrections to R$_{90}$ are negligible as R$_{90}$ is well resolved.

Despite the different nebulae sizes, most of the galaxies have a remarkably similar \oii \ R$_{50}$, with a median value of 6 kpc. J0826 and J1613 have substantially larger \oii \ nebulae with R$_{50}$ $\sim$10.5 kpc. Comparing R$_{50}$ with the maximum radial extents reported in Section~\ref{sec:mor}, we note that in our sample on average 50\% of the \oii \ light is contained within $\sim$1/3 of the maximum radial extent of the nebula, corresponding to $\sim$15 percent of the total covered area. Interestingly, R$_{90}$ is also substantially smaller than the maximum radial extents, indicating that the \oii \ gas is less luminous in the outskirts of the nebulae and traces less dense material at large distances.

The two most extended \oii \ nebulae in our sample, J0826 and J1613, are also the most asymmetric, which can be seen in their SB radial profiles as a bump at large radii. In the case of J0826, the SB bump is due to the gas concentrated around a prominent tidal tail extending to the southeast of the galaxy, while for J1613 it corresponds to the gas around a small galaxy $\sim$ 40 kpc away with which it is interacting \citep{sel14}.

To take into account the substantial asymmetry observed in these two \oii \ nebulae, we calculate R$_{50}$ and R$_{90}$ again, now collapsing their SB maps along the two axes in which the nebulae are elongated. We rotate the J0826 SB map 20 degrees counter-clockwise to have the axis of maximum extent corresponding to a vertical axis. We collapse the SB map along the vertical and horizontal axes, measuring R$_{50}$ and R$_{90}$ along each, obtaining 6 and 24 kpc, and 4 and 10 kpc, respectively. For J1613, we decide to consider only the top part of the nebula (down to $\sim -25$ kpc) and exclude the bottom lobe, which is due to an interaction with the nearby galaxy. We collapse the SB map along the north-south and east-west axes, obtaining very similar R$_{50}$ of 3.5 kpc, and R$_{90}$ of 13.9 kpc and 15.6 kpc, respectively. We note that in both cases, the R$_{50}$ measured with this different approach better captures the central concentration of the \oii \ gas distribution and is more in line with the rest of the galaxy sample.

\begin{deluxetable*}{lrCCcDD}
\tablecaption{\mgii\ Properties\label{table4}}
\tablehead{
\colhead{Object Name } & \colhead{Area}  & \colhead{L(\mgii)} & \colhead{L(\mgii)$_{dust}$} & \colhead{EW} & \twocolhead{R$_{50}$} & \twocolhead{$\frac{\rm R_{50}(MgII)}{\rm R_{50}([OII])}$} \\
\colhead{}  &\colhead{(kpc$^2$)}  & \colhead{(erg s$^{-1}$)}& \colhead{(erg s$^{-1}$)} & \colhead{(\AA)} & \twocolhead{(kpc)} & \twocolhead{}\\
\colhead{(1)} & \colhead{(2)} & \colhead{(3)} & \colhead{(4)} & \colhead{(5)} & \twocolhead{(6)} & \twocolhead{(7)}  }
\decimals
\startdata
J0826+4305    &  476 & 2.7\times 10^{41} & \nodata           & 0.97 & 10.0 & 0.94 \\
J0944+0930    &  292 & 6.4\times 10^{40} & 3.8\times 10^{41} & 0.55 &  4.9 & 0.87 \\
J1107+0417    &  367 & 1.7\times 10^{41} & 5.4\times 10^{41} & 1.01 &  5.1 & 0.95 \\
J1205+1818    &  289 & 6.2\times 10^{41} & 9.4\times 10^{42} & 0.73 &  3.9 & 0.72 \\
J1244+4140    &  387 & 8.4\times 10^{40} & \nodata           & 0.76 &  5.3 & 0.86 \\
J1341$-$0321  &  458 & 4.3\times 10^{41} & 1.2\times 10^{42} & 0.85 &  6.3 & 1.00 \\
J1500+1739    &  417 & 4.9\times 10^{41} & 2.8\times 10^{42} & 2.53 &  4.0 & 0.72 \\
J1506+5402    &  819 & 5.5\times 10^{41} & 1.0\times 10^{42} & 1.76 &  5.4 & 0.69 \\
J1558+3957    &  256 & 3.3\times 10^{40} & 8.4\times 10^{40} & 0.31 &  3.3 & 0.42 \\
J1613+2834    &  414 & 1.6\times 10^{41} & 4.8\times 10^{41} & 1.50 &  5.2 & 0.50 \\
J1622+3145    &  519 & 3.3\times 10^{41} & 8.6\times 10^{41} & 2.52 &  5.6 & 0.74 \\
J1713+2817\tablenotemark{a} & \nodata & 1.3\times 10^{41} & 6.3\times 10^{41} & 1.30 &  \nodata & \nodata \\
\hline
J2118+0017    &  600 & 2.3\times 10^{41} & 3.0\times 10^{42}\tablenotemark{b} & 3.51 & 9 & 0.5
\enddata
\tablecomments{Col 2: Area of the 1$\sigma$ \mgii\  nebula; Col 3: \mgii \ luminosity corrected for Galactic extinction; Col 4: \mgii \  luminosity corrected for Galactic and galaxy intrinsic dust extinction; Col 5: rest-frame \mgii \ equivalent width; Col 6: radius containing 50\% of the light in the surface brightness radial profile; Col 7: \mgii \ to \oii \ R$_{50}$ ratio, using Col (6) and Table~\ref{table3} Col (6).}
\tablenotetext{a}{The \mgii \ signal per spaxel in J1713 is too faint to produce a reliable \mgii \ SB map and measure the area or $R_{50}$.}
\tablenotetext{b}{See footnote to Table~\ref{table3} for extinction model.}
\end{deluxetable*}

\begin{figure*}[htp!]
 \centering
 \includegraphics[width=0.85\textwidth]{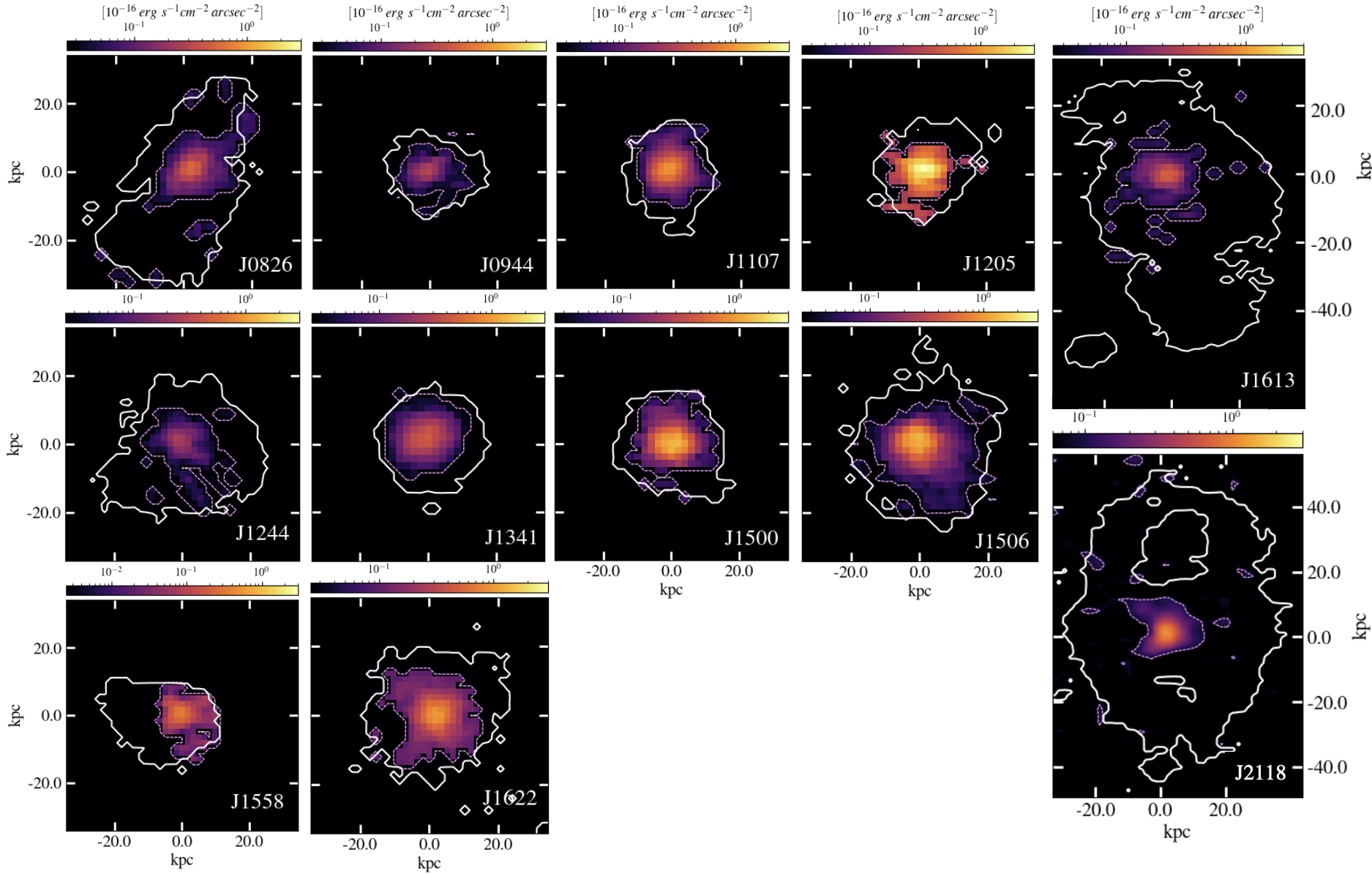}
 \caption{Colors show observed-frame \mgii \ surface brightness for the galaxies in our sample. On each panel, we show the area with a signal that looks real above a threshold defined after an eye inspection of the extracted spectra. The dashed white contour denotes 1$\sigma$ \mgii \ emission, while the solid white one shows the 4$\sigma$ \oii \ emission as presented in Figure~\ref{fig:o2_sb}. The axes are labeled in kpc from the brightest spaxel. The \mgii \ signal per spaxel in J1713 is too faint to produce a reliable \mgii \ SB map.}
 \label{fig:mg2_sb}
\end{figure*}

\subsection{MgII emission}\label{sec:eml}

Here we study the spatial distribution of the emission arising from the \mgii \ transition.

As \mgii \ is a resonant line the emitted photons are constantly reabsorbed due to the absence of fine structure splitting. This trapping mechanism interferes with the escape of the photons, complicating the interpretation of the origin of \mgii \ emission. In the traditional model of a galactic-scale outflow expanding as a shell, this produces a P-Cygni-like profile for each \mgii \ doublet component, with blueshifted absorption and redshifted emission.  

The galaxies in our sample commonly display extreme kinematics in the \mgii \ absorption lines that have been extensively studied in \citet{per23}. Another common characteristic in our sample is faint \mgii \ emission. In most cases, we observe only one of the emission doublet components, \mgii$\lambda$2803, as the corresponding \mgii$\lambda$2796 line is not visible due to \mgii$\lambda$2803 absorption at the same wavelength. The continuum-subtracted \mgii \ spectral regions for the galaxies in our sample are included in Appendix~\ref{app:A} and illustrate the range of \mgii \ emission and absorption properties in our data.

Figure~\ref{fig:mg2_sb} shows the \mgii \ SB maps. In each panel the white dashed contour corresponds to the \mgii \ 1$\sigma$ detection significance, corresponding to an average surface brightness limit of 2.5 $\times$ 10$^{-18} \, erg \,s^{-1} \, cm^{-2} \, arcsec^{-2}$. The low \mgii \ signal-to-noise ratio per spaxel prevents us from utilizing a 4$\sigma$ significance threshold as for \oii, as the results maps would include only a few spaxels at the center of the nebulae. Therefore, we intentionally use a lower significance level for \mgii, aware that this results in inflating the extent of \mgii \ compared to \oii. To ease comparison between the \mgii \ and \oii \ maps we display the 4$\sigma$ \oii \ emission level as a white solid contour, as illustrated in Figure~\ref{fig:o2_sb}. 

Even using these different significance thresholds, the \mgii \ emission is substantially less extended than \oii \ in all galaxies in our sample. We report in Table~\ref{table4} the area covered by the \mgii \ nebulae applying a 1$\sigma$ S/N cut. The 1$\sigma$ \mgii \ emission covers on average only $\sim$ 43 percent of the area covered by the 4$\sigma$ \oii \ emission, which indicates that the \mgii \ extent is far smaller than that of \oii. The \mgii \ emission observed in Makani is a factor of 2 smaller than the \oii\ emission, in line with our sample. An important way in which Makani differs from the other compact starburst in this work is the absence of strong and highly blueshifted \mgii \ absorption. Makani exhibits no visible absorption in its integrated spectrum, though some is present . We discuss the possible reason for such a difference in Section~\ref{sec:discussion} below.

We produce SB radial profiles for \mgii \ as described in Section~\ref{em_mg_sb}. The SB profiles are noisy, particularly at large radial distances. Therefore we calculate only R$_{50}$, as the large uncertainties result in unreliable R$_{90}$ estimates. The values of the \mgii \ R$_{50}$ deconvolved with the seeing for each galaxy are reported in Table~\ref{table4}. 

In Table~\ref{table4} we also report the \mgii \ to \oii \ R$_{50}$ ratio, which has a median value of 0.75. This shows that even the inner light profiles of \mgii \ are less extended than that of \oii,  with \oii \ exhibiting on average $\sim$ 33\% larger radial extent. If we were able to obtain a reliable measurement of  R$_{90}$(\mgii), it would be substantially smaller than R$_{90}$(\oii), implying that \oii \ can be used to trace ionized gas farther from the center of the galaxy.

Unlike \oii, \mgii \ emission is not detected in spatial correspondence with stellar emission from merger debris. This suggests that the two ions can trace different physical mechanisms or have different origins. The most evident cases of this in our sample are J0826, J1244, J1558, and J1613. We return to this point in Section~\ref{sec:discussion} below.

\subsection{Kinematics}\label{sec:kin}

Integral field spectral data allow us to produce spatially-resolved kinematics maps of the ionized gas in our galaxies. We quantify the \oii \ kinematics from the line profile fits. At each spaxel we measure the line velocity dispersion ($\sigma$) and the velocity shifts relative to the systemic redshift (determined by the central velocity of the stellar continuum) at which 2\% ($v_{02}$), 50\% ($v_{50}$), and 98\% ($v_{98}$) of the line flux accumulates integrating from red (positive velocities) to blue (negative velocities) across the line profile. $v_{50}$ is the median velocity in the profile, while $v_{02}$ and $v_{98}$ measure the maximum redshifted and blueshifted wings, respectively. We report $v_{50}$, $v_{02}$, $v_{98}$, and $\sigma$ measured across each nebula in our sample in Table~\ref{table3}, where the median and maximum values are calculated after clipping the velocity distributions at 3$\sigma$ to eliminate outliers.

For the following discussion, we assemble the nebulae into three groups based on kinematics and morphology. 

\subsubsection{Highest velocity dispersion}

\begin{figure*}[htp!]
 \centering
  \includegraphics[width=0.9\textwidth]{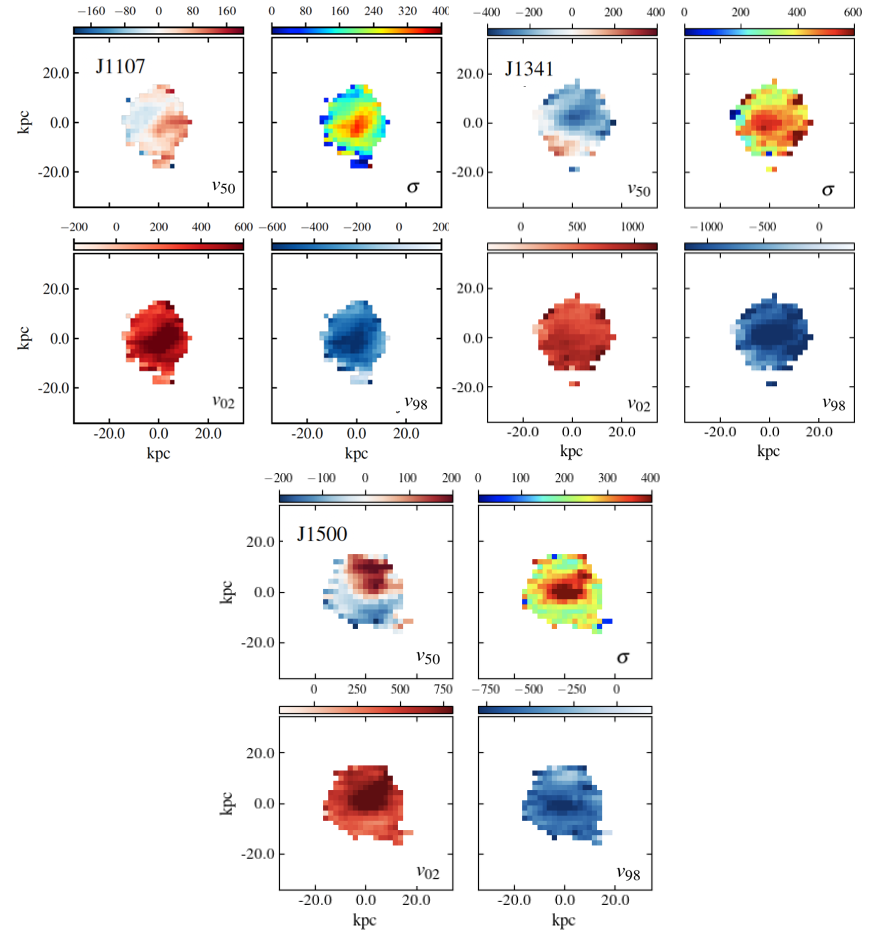}

 \caption{ \oii \ kinematics for three of the galaxies in our sample with the highest values of velocity dispersion, $\sigma$, J1107 (top left four panels), J1341 (top right four panels), and J1500 (bottom four panels). In each set of four panels, we show in the top left panel the \oii \ central velocity ($v_{50}$); in the top right panel the \oii \ velocity dispersion ($\sigma$), in the bottom left panel the \oii \ maximum redshifted velocity ($v_{02}$); in the bottom right panel \oii \ maximum blueshifted velocity ($v_{98}$). All velocities of the ionized gas are relative to the systemic redshift of the source. The color bars show the velocity in \kmps. The axes are labeled in kpc from the brightest spaxel.}
 \label{fig:quad_1}
\end{figure*}

Figure~\ref{fig:quad_1} shows the kinematics maps for three of the four nebulae in our sample with the highest velocity dispersion.
In each set of four panels, the upper left panel displays the central velocity, $v_{50}$, for that nebula. The $v_{50}$ maps reveal asymmetric velocity gradients across the nebulae in J1107 and J1500, while J1341 is mostly blueshifted.
 
\begin{figure*}[]
 \centering
 \includegraphics[width=0.9\textwidth]{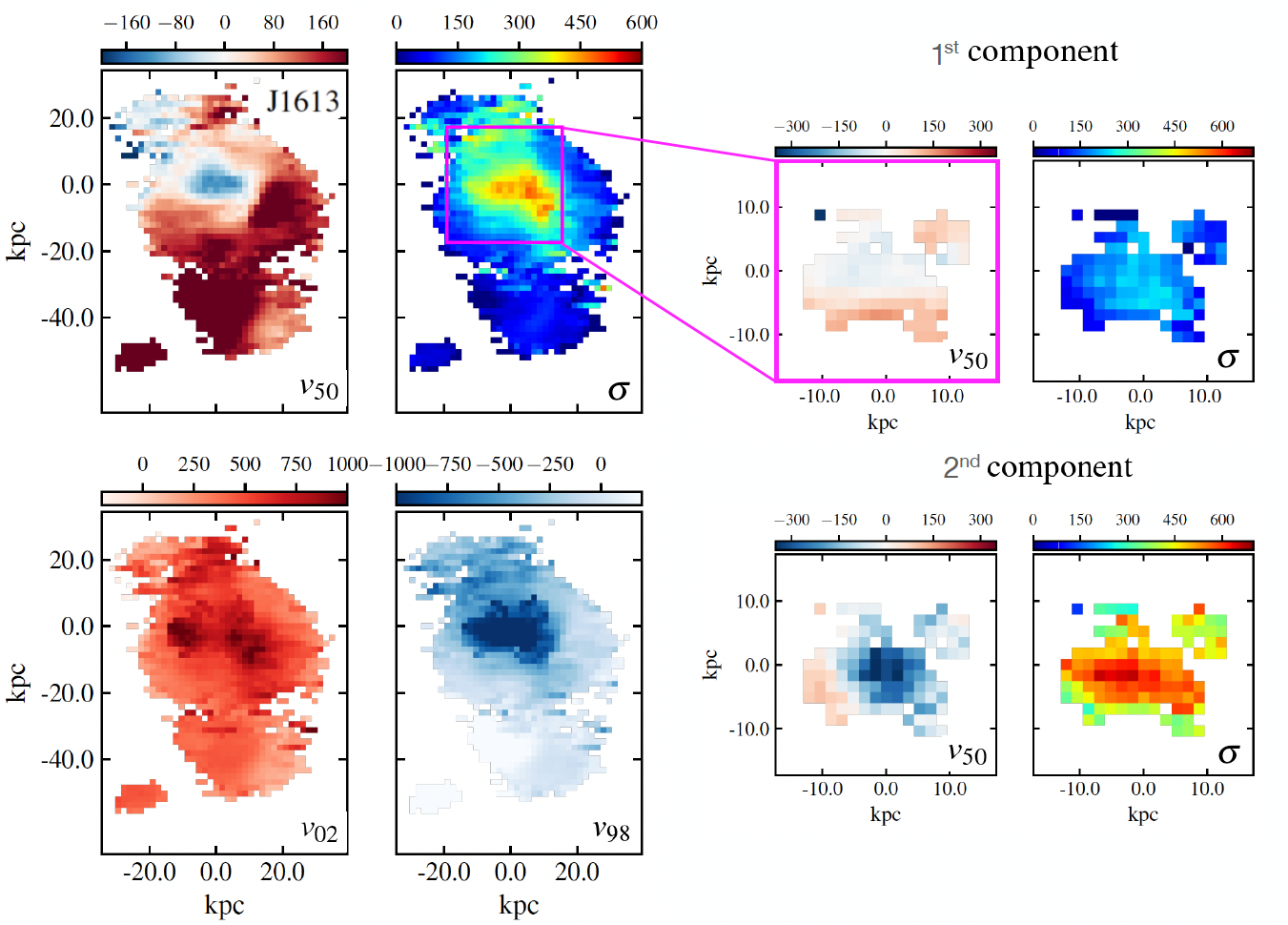}

 \caption{Kinematics of the \oii \ emission for the galaxy J1613. The axes are labeled in kpc from the brightest spaxel. Left: The four panels show the \oii \ velocity ($v_{50}$, $v_{02}$, and $v_{98}$) and velocity dispersion ($\sigma$) maps  for the galaxy J1613. The magenta box, shown in the remaining right panels, traces the central $6'' \times 6''$. Right: The four panels show the $v_{50}$ and $\sigma$ maps of the \oii \ emission for the first (top) and second (bottom) \oii \ kinematic components.}
 \label{fig:1613_kin}
\end{figure*}

The upper right panels of each set of four panels in Figure~\ref{fig:quad_1} show the velocity dispersion, $\sigma$, across each nebula. The central regions of J1107, J1341, and J1500 have very broad emission with $\sigma \sim$ 300, 500, and 400 \kmps, respectively. 

In each set of four panels of Figure~\ref{fig:quad_1} the lower left and right ones show the maximum redshifted and blueshifted velocities $v_{02}$ and $v_{98}$. The three nebulae have fairly regular $v_{02}$ and v$v_{98}$ maps, with peak values corresponding to the regions of highest velocity dispersion, as expected given the broader line profiles in those regions. 

The fourth system in our sample with the highest velocity dispersion is J1613 (Figure~\ref{fig:1613_kin}). J1613 is interacting with a smaller galaxy $\sim$40 kpc to the south, at almost the same redshift. Its \oii \ nebula has three separate regions with distinct kinematics signatures: 1) the nuclear region surrounding the compact starburst and extending up to $\sim15$ kpc north and $\sim25$ kpc south, 2) the northern lobe surrounding an evacuated region centered at $\sim$21 kpc northeast of the center of the galaxy, and 3) the southern lobe overlapping with the region of interaction with a nearby galaxy located $\sim$40 kpc south. There is also a third small galaxy in the J1613 field at the same redshift, located $\sim$52 kpc southeast.  The \oii \ nebula does not display contiguous emission connected to this third galaxy.

The nuclear region exhibits complex kinematics and at the center requires a second Gaussian component to fit the \oii \ emission line profiles accurately. 
The magenta box in the $\sigma$ map panel marks the central $6'' \times 6''$, which are shown in the right panels of Figure~\ref{fig:1613_kin} where the top and bottom rows display the \oii \ $v_{50}$ and $\sigma$ measured from the fit of the first and second kinematic components, respectively. 
The left panels in this figure, showing the total emission, reveal that the center of the nuclear region has a mild blueshift of $\sim60 - 130$ \kmps ~and very broad emission with an average $\sigma \sim450$ \kmps. This blueshift and large $\sigma$ are driven by the second kinematic component in this region, which shows a median $v_{50}$ of $-$110 \kmps ~and peak $v_{50}$ values of $-$400 \kmps, and a median $\sigma$ of 500 \kmps ~and peak values of 650 \kmps (see the bottom right panels). The $v_{50}$ of the first kinematic component exhibits very mild blueshifts and redshifts with a median value close to systemic. The $\sigma$ of the first kinematic component reveals relatively narrow emission with a median $\sigma$ of 170 \kmps.
The rest of the nuclear region surrounding the central blueshift is mostly redshifted, ranging from a few tens of \kmps ~to peaks of $\sim$300 \kmps. This region also exhibits a decreasing $\sigma$ in the map showing the total emission, moving radially out from the center of the galaxy.
Notably, some of the most redshifted gas in the $v_{50}$ map lies at the edge of the limb-brightened bubble surrounding the evacuated region $\sim$16 kpc southwest of the central starburst, seen in Figure~\ref{fig:o2_sb}. 

The northern lobe of the \oii \ nebula in J1613 exhibits an east-west $v_{50}$ gradient with blueshifted and redshifted peak values of $\sim$200 \kmps. The $\sigma$ map in this region  
reveals that most blueshifted gas has lower line widths with $\sigma \sim100 - 150$ \kmps, while most redshifted gas has higher $\sigma \sim200 - 350$ \kmps.

The southern lobe of the \oii \ nebula is redshifted by $\sim40 - 400$ \kmps. The $\sigma$ map in this region is quite uniform and shows narrow emission with $\sigma <100$ \kmps ~and a median value of $\sim$50 \kmps. The gradient across this lobe could trace disk rotation in this galaxy.

The $v_{02}$ and $v_{98}$ maps show that the most blueshifted and redshifted gas is near the center of the nuclear region and is due to the presence of the second kinematic component.
We discuss more the presence of the second kinematics component in Section~\ref{sec:discussion}.

\begin{figure*}[htp!]
 \centering
 \includegraphics[width=0.9\textwidth]{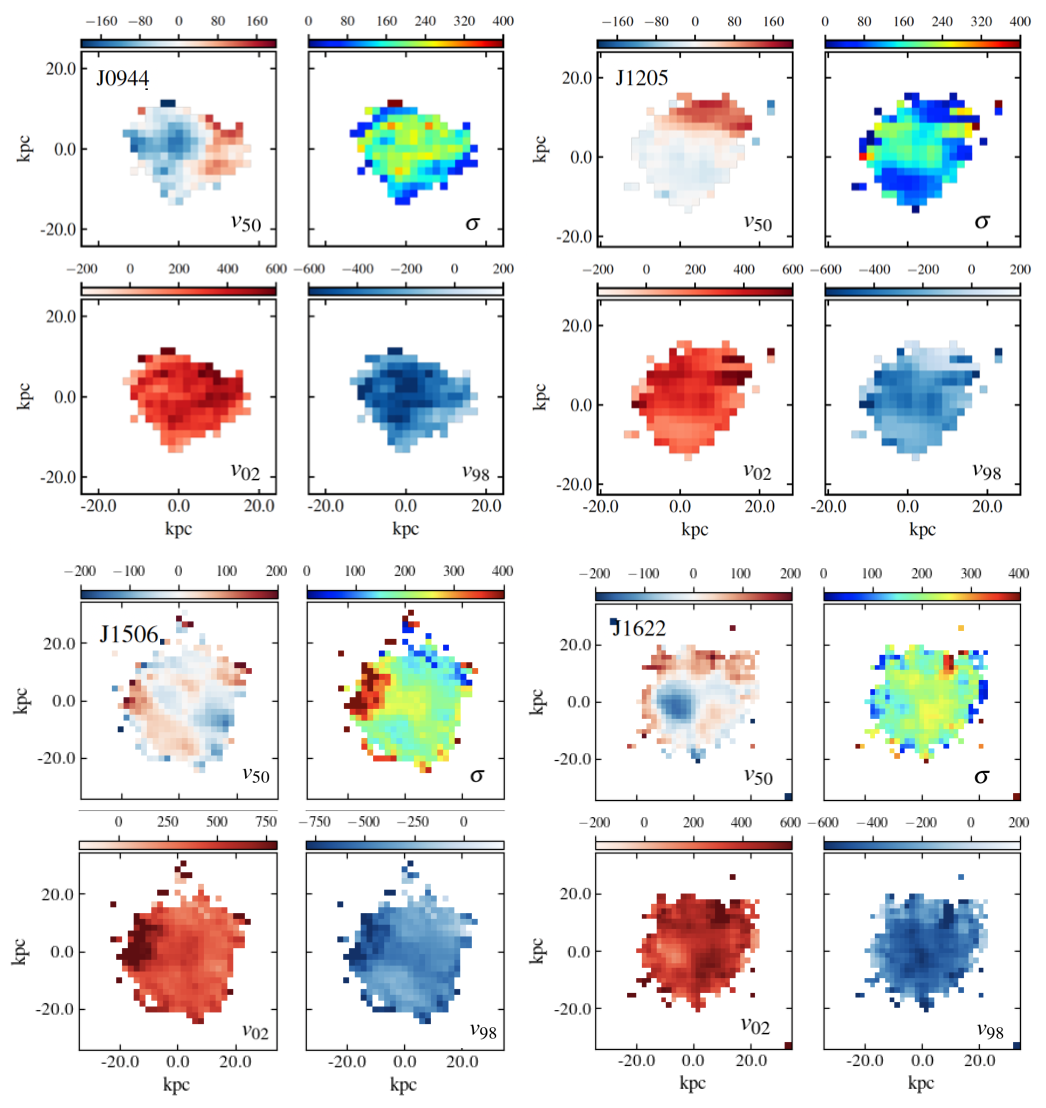}
 \caption{ \oii \ kinematics for four galaxies in our sample with moderately high velocity dispersion, $\sigma$, J0944, J1205, J1622, and J1506.  In each quadrant, we show in the top left panel the \oii \ central velocity ($v_{50}$); in the top right panel the \oii \ velocity dispersion ($\sigma$), in the bottom left panel the \oii \ maximum redshifted velocity ($v_{02}$); in the bottom right panel \oii \ maximum blueshifted velocity ($v_{98}$). All velocities of the ionized gas are relative to the systemic redshift of the source. The color bars show the velocity in \kmps. The axes are labeled in kpc from the brightest spaxel.}
 \label{fig:quad_2}
\end{figure*}

\begin{figure}[htp!]
 \centering
 \includegraphics[width=0.4\textwidth]{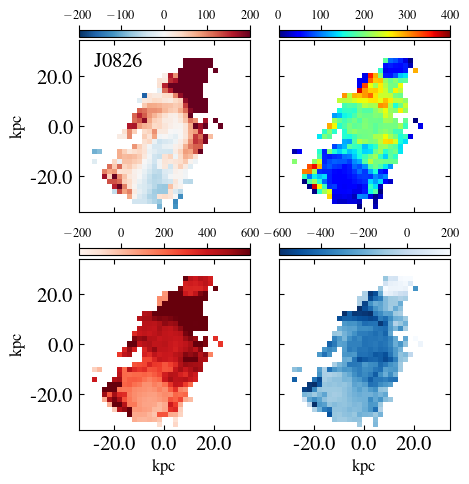}

 \caption{\oii \ kinematics for the galaxy J0826. All velocities of the ionized gas are relative to the systemic redshift of the source. We show in the top left panel the \oii \ central velocity ($v_{50}$); in the top right panel the \oii \ velocity dispersion ($\sigma$), in the bottom left panel the \oii \ maximum redshifted velocity ($v_{02}$); in the bottom right panel \oii \ maximum blueshifted velocity ($v_{98}$). All velocities of the ionized gas are relative to the systemic redshift of the source. The color bars show the velocity in \kmps. The axes are labeled in kpc from the brightest spaxel.}
 \label{fig:0826_kin}
\end{figure}

\subsubsection{Moderate velocity dispersion}

Figure~\ref{fig:quad_2} presents kinematics maps for four galaxies in our sample with moderate velocity dispersion. 
The $v_{50}$ maps show asymmetric velocity gradients across J0944 and J1205. J1506 shows a complex, irregular $v_{50}$ map with several regions with small blueshifts or redshifts. The $v_{50}$ map in J1622 exhibits moderate blueshifts and redshifts. The largest redshifts are at the north and west edges of the nebula. The maximum blueshifted $v_{50}$ values are concentrated in a circular region $\sim$7 kpc to the east of the galaxy's center. 

The central regions of J0944, J1205, and J1622 show moderately broad emission with peaks of $\sigma \sim$300, 280, and 250 \kmps, respectively. The $\sigma$ values are lower in the outskirts of these nebulae, as low as $\sim$50 \kmps. J1506 has a substantially different $\sigma$ map than the other three galaxies in this group. In the center, the emission is broad, with $\sigma \sim$240 \kmps, but the region with the highest velocity dispersion of $\sigma >$350 \kmps~is off-center and lies at the northeastern edge of the nebula. 
Interestingly, the J1622 $\sigma$ is lower than the values in the center in correspondence with the highly blueshifted gas. Furthermore, the highest $\sigma$ values are located in a small region at the northwest edge of the nebula, with peaks of $\sigma \sim$400 \kmps. The line profile fit in this small region requires a second broad kinematic component.

The fifth galaxy with moderate velocity dispersion is J0826 (Figure~\ref{fig:0826_kin}). This galaxy exhibits a complex kinematic structure. There are three regions with distinct kinematic features: 1) the nuclear region around the compact starburst; 2) the large-scale gas extending around and beyond a prominent tidal feature to the southeast of the galaxy; and 3) the large-scale gas extending $\sim$30 kpc to the northwest of the galaxy. The nuclear region shows uniformly broad emission with average $\sigma \sim200$ \kmps\ and a mild velocity gradient. The narrowest line widths are found around the tidal tail, where the gas has an average $\sigma \sim$ 50 \kmps. 
The extended gas to the northwest is redshifted and the $\sigma$ map in this region is complex, with the lowest values, $\sigma \sim$ 50 \kmps, at the northwestern edge of the nebula. The broadest line widths in the nebula are found $\sim$14 kpc north of the central starburst, at the base of the northwest region, with values up to 350 \kmps. 

\subsubsection{Low velocity dispersion}

\begin{figure*}[htp!]
 \centering
 \includegraphics[width=0.9\textwidth]{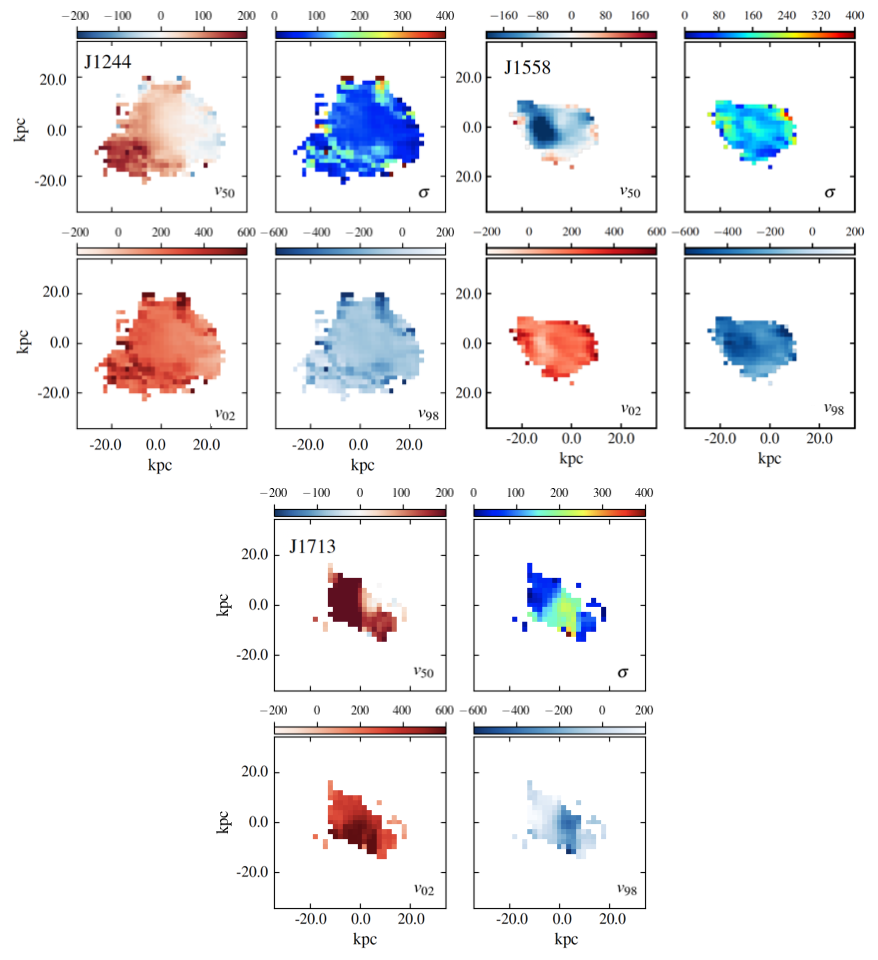}
 \caption{\oii \ kinematics for the three galaxies in our sample with the lowest velocity dispersion in our sample J1244 (top left four panels), J1558 (top right four panels), and J1713 (bottom four panels). In every four panels, we show in the top left the \oii \ central velocity ($v_{50}$); in the top right the \oii \ velocity dispersion ($\sigma$), in the bottom left the \oii \ maximum redshifted velocity ($v_{02}$); in the bottom right \oii \ maximum blueshifted velocity ($v_{98}$). All velocities of the ionized gas are relative to the systemic redshift of the source. The color bars show the velocity in \kmps. The axes are labeled in kpc from the brightest spaxel.}
 \label{fig:duo}
\end{figure*}

Figure~\ref{fig:duo} shows the kinematics maps for the three galaxies in our sample with the lowest velocity dispersion. 

The top left panels illustrate the kinematics for J1244.
The $v_{50}$ map shows asymmetric velocity gradients across J1244. Most J1244 $\sigma$ map exhibits relatively narrow emission line profiles with median $\sigma \sim$75 \kmps.
The evident tidal tail extending to the southwest of J1244, seen in Figure~\ref{fig:fig2}, overlaps with the smallest \oii \ $\sigma$ with values as low as $\sim$ 30 \kmps.

The $v_{50}$ map in J1558 shows mostly blueshifted emission. As observed in J1622, the maximum blueshifted $v_{50}$ values in J1558 are concentrated in a slightly elongated region off-center 8 kpc to the west. J1558 is an ongoing merger and the maximum blueshifted $v_{50}$ region lies atop these tidal features. It has an irregular $\sigma$ map, with several regions with moderately broad emission. The ongoing merger event may be causing some of the disturbed kinematics observed. The gradient in the center could potentially trace a compact, rotating disk.

J1713 has mostly redshifted gas. However, it also possesses an unresolved, high-velocity outflow. Aside from this, J1713 exhibits moderately broad emission across its central region, with peak values of $\sigma \sim$230 \kmps, and narrow line profiles at either end of the nebula. This source is a close binary in projection with small-scale tidal features, as revealed at {\it HST} resolution (Figure~\ref{fig:fig2}). The disjoint kinematics of the gas and stars are likely a reflection of this close interaction. J1713 is also a type II AGN candidate. Finally, it has a companion galaxy 54~kpc (in projection) to the NW, which we confirm to be at the same redshift.

\begin{figure*}[htp!]
 \centering
 \includegraphics[width=0.9\textwidth]{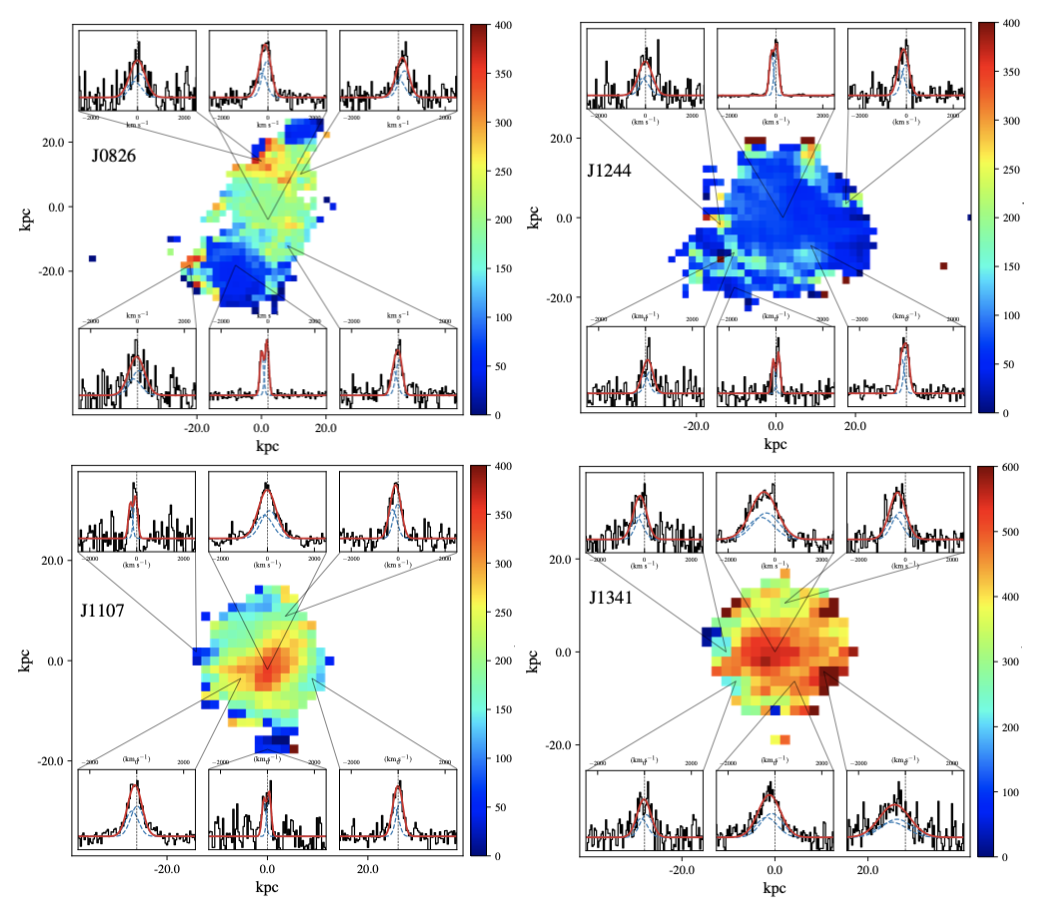}
 \caption{Velocity dispersion $\sigma$ map of the \oii \ emission for J0826, J1107, J1244, and J1341. Velocity profiles of representative spaxels are shown in the insets, highlighting areas with high and low velocity dispersion. In the velocity profiles and spectra, the black line is the continuum-subtracted spectrum, the red solid line is the total emission line model, and the blue dashed lines are the \oii \ emission line models for the individual doublet components.
 }
 \label{fig:inset1}
\end{figure*}

\begin{figure*}[htp!]
 \centering
 \includegraphics[width=0.9\textwidth]{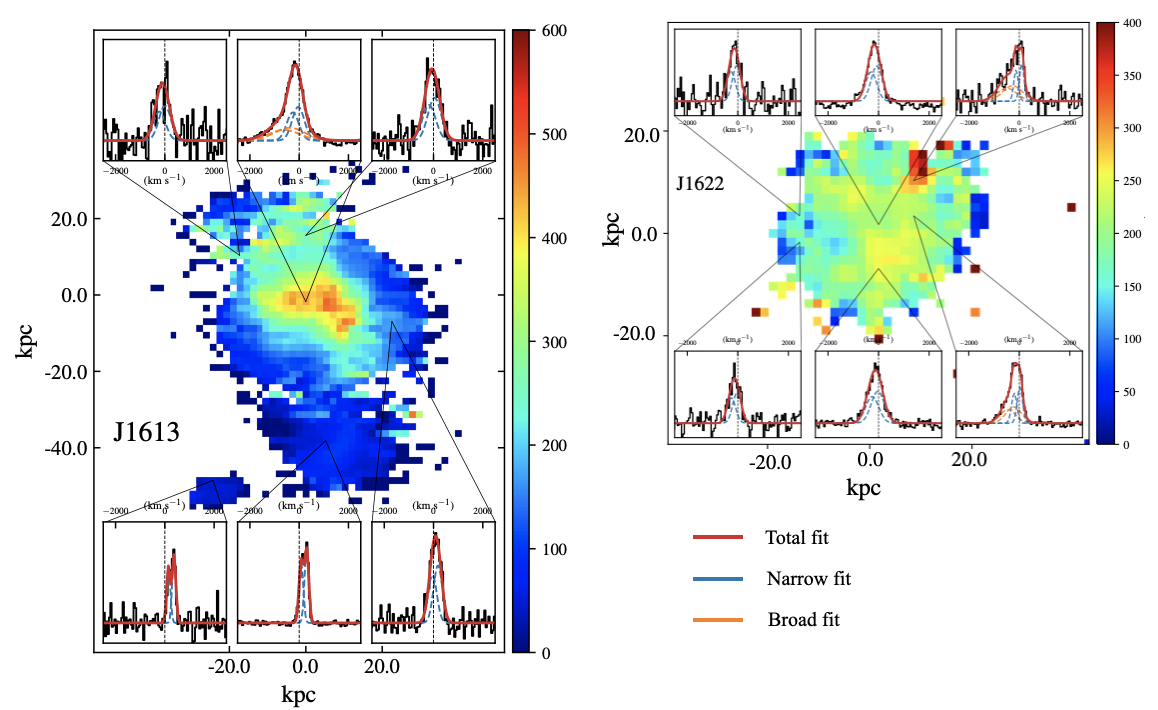}
\caption{Velocity dispersion $\sigma$ map of the \oii \ emission for J1613 and J1622. See Figure~\ref{fig:inset1} for more details. In addition, orange dashed lines are the emission line models for the broad kinematic components.}
 \label{fig:inset2}
\end{figure*}

Figures~\ref{fig:inset1} and \ref{fig:inset2} show the quality of the \oii \ emission line spectra and best-fit models in various locations for six of the galaxies in our sample. Examples of spectral fits for the remaining six galaxies are included in Appendix~\ref{app:A}. Each panel of Figures~\ref{fig:inset1} and ~\ref{fig:inset2} displays a $\sigma$ map with velocity profiles of representative spaxels in the insets, highlighting areas with either high or low velocity dispersion. We note that a second kinematic component is required in the central region of J1613 and in the northwestern edge of J1622. 

\section{Discussion}\label{sec:discussion}

We detect emission nebulae in 12 compact, massive starburst galaxies tracing \oii \ and \mgii, which allows us to probe the morphology and kinematics of the ionized gas in and around these galaxies to large distances. We show that \oii \ is a better tracer than \mgii \ of the low surface brightness, extended emission. The \oii \ nebulae in our sample are spatially extended beyond the stars in these galaxies, with radial extent, R$_{90}$, ranging between 10 and 40 kpc, and maximum blueshifted speed, $v_{98}$, ranging from -335 to -1920 \kmps ~(Table~\ref{table3}). 

We now interpret the \oii \ emission in our sample as galactic outflows (Section~\ref{sub:outflows}) and discuss a possible relation between the \oii\ nebula and star formation history (SFH) of each galaxy (Section~\ref{sub:sfh_link}). In Section~\ref{sub:mout}, we address the differences between estimating mass outflow rates using emission and absorption lines. We conclude with a brief discussion of the implications for starburst-driven feedback and how the HizEA sample might fit into the merger-driven galaxy evolutionary scenario (Section~\ref{sub:implications}).

 \begin{figure*}[htp!]
 \centering
 \includegraphics[width=0.9\textwidth]{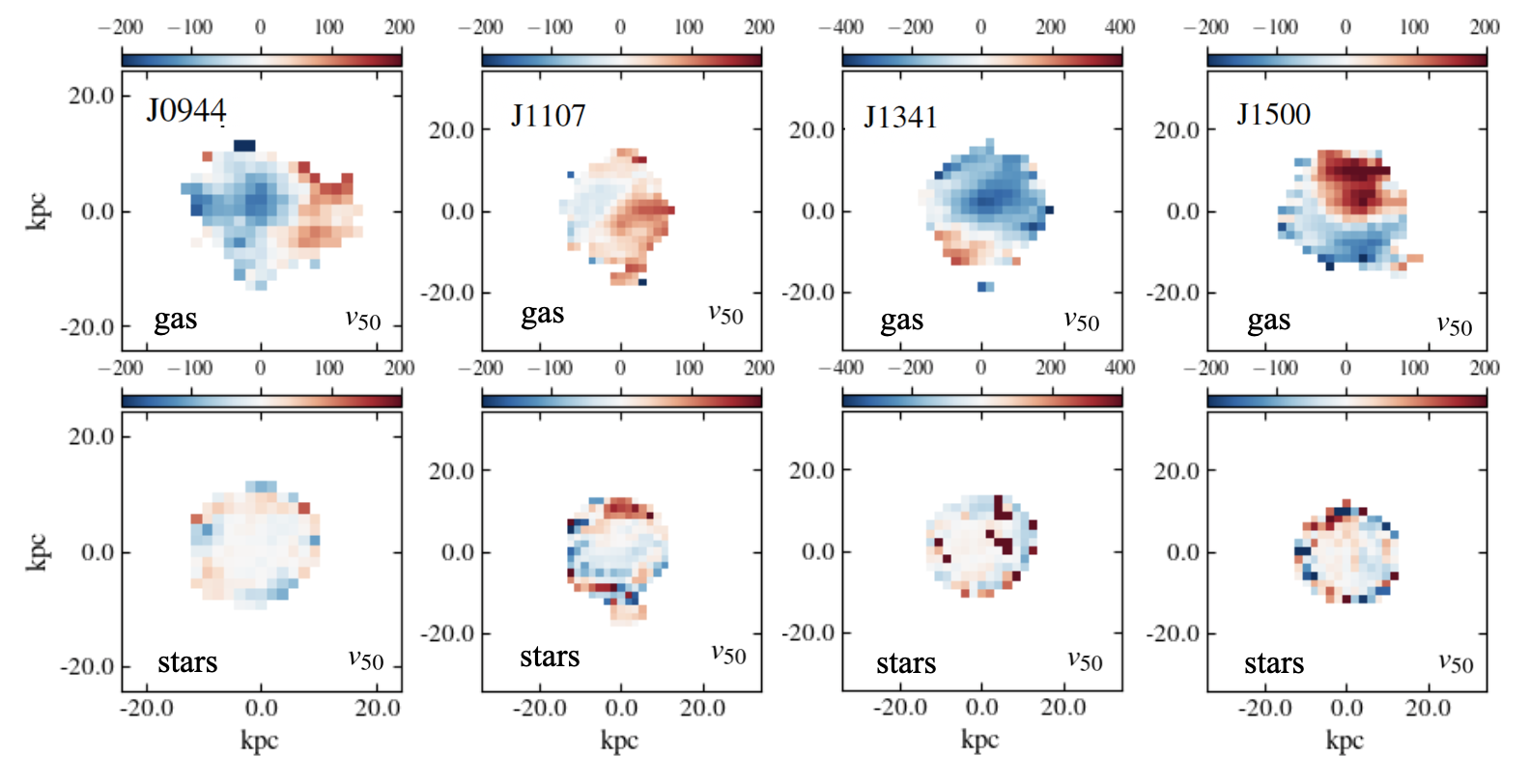}
 \caption{\oii \ and stellar kinematics for four galaxies in our sample that show an \oii \ emission velocity gradient J0944 (left two panels), J1107 (second two panels), J1341 (third two panels) and J1500 (right two panels). In every two panels, we show at the top the \oii \ central velocity ($v_{50}$) and at the bottom the stellar $v_{50}$. All velocities of the ionized gas are relative to the systemic redshift of the source. The color bars show the velocity in \kmps. The axes are labeled in kpc from the brightest spaxel.}
 \label{fig:kin_gas_star}
\end{figure*}

\subsection{\oii \ Emission Tracing Spatially-Resolved Outflows}\label{sub:outflows}

Galactic winds are typically identified through their physical extents and kinematic signatures. As outflows evacuate gas from star formation regions into the CGM, this gas moves beyond the galaxy stellar continua \citep[e.g.][]{har12, ho16, rup19, zab20, dut23}. As we showed in Section~\ref{sec:mor} and \ref{sec:rad_profile}, the \oii \ emission is more extended than the stellar continuum in all of the galaxies of our sample. We note that the observed size of the \oii \ nebulae (Figure~\ref{fig:o2_sb}) is a lower limit on the true extent, as we mask out spaxels with S/N $<$ 4. Moreover, emission lines are sensitive to the density squared of the gas probed, which results in missing lower-density, weaker gas components that can be more diffuse and extended. 

In addition to size compared to the stellar continuum, the identification of outflowing gas---in cases where the outflow is spatially resolved---often relies on a comparison between the observed gas and star velocity fields and line width distributions. 

In Section~\ref{sec:kin}, we showed that half of the galaxies in our sample do not have an ordered rotation field in \oii. Additionally, those that do exhibit a velocity gradient in \oii\ have high $\sigma$ values, in which a substantial portion of the gas is moving with velocities of several hundreds of \kmps ~(see Table~\ref{table3}). This is inconsistent with this gas being in dynamical equilibrium with the host galaxy or with simple disk rotation.

For each galaxy in our sample, we model the stellar continuum in each spaxel (Section~\ref{subsec:emlfit}) and produce stellar $v_{50}$ maps to compare to those of \oii. Figure~\ref{fig:kin_gas_star} shows the \oii \ and stellar kinematics maps for four galaxies in our sample with a velocity gradient in their \oii \ $v_{50}$ maps. This comparison verifies that the \oii \ emission does not reflect rotation with the same velocity gradient seen in the stars. This conclusion also applies to the rest of the galaxies in our sample. The HizEA galaxies are not disk galaxies, as revealed by HST imaging \citep{dia12, sel14}. We confirm that they are dispersion-dominated systems, as the $v_{50}$ maps of the stars show no (or only weak) velocity gradients. The ratio of the stellar $v_{50}$ to $\sigma$ in our sample is low, with a median value of 0.23; this is an order of magnitude lower than what is seen in star-forming galaxies \citep[e.g.][]{gre14}. Given the \oii \ emission extends beyond the stellar continuum and that the \oii \ kinematics are inconsistent with dynamical equilibrium with the host galaxy or disk rotation, we conclude that the \oii \ nebulae in our sample are dominated by non-gravitational motions.

Previous studies \citep{rup19, per23} have revealed that the galaxies in our sample exhibit episodic outflows. In particular, as recapitulated in Section~\ref{makani}, the outflowing wind in Maknai cleanly separates into two episodes with distinct extents and kinematics. In this regard, J1613 is similar to Makani as it requires a second kinematic component to fit the \oii \ emission line profile in the central $\sim$ 20 kpc around the compact starburst (see Figure~\ref{fig:1613_kin}). The second component is broad and traces a kinematically distinct part of the outflowing gas. 

While we do not have sufficient signal-to-noise per spaxel for the rest of our sample to identify a second kinematic component in spatially-resolved maps, 9 of the 12 galaxies do exhibit an additional broad component in spatially-integrated spectra (Figures~\ref{fig:integrated}-\ref{fig:int_3}). The median values of the \oii \ $\sigma$ broad components in the full nebula and core spectra across our sample are 560 and 670 \kmps, respectively. The broad components are offset in their centroid velocities, $v_{50}$, from the systemic velocity, blueshifted by median values of 225 and 155 \kmps\ in the full nebula and core spectra, respectively. The best-fit parameters obtained from the \oii \ core spectra are in good agreement with our previous long-slit data \citep{per21}. Such line broadening and blueshifts are typically interpreted as outflowing gas. For most galaxies in our sample the broad components include some redshifted emission as well compared to the systemic, though the velocity centroids are always blueshifted. We attribute this to dust present in the host galaxy that obscures a portion of the redshifted outflows. Lastly, we note that the broad to narrow flux ratio of the \oii \ line profile integrated over the whole nebula is larger than the ratio observed in the core for all galaxies where we observe a broad component. This result suggests that the broad component traces outflowing gas more extended than the central 5$\times$5 spaxels.

For most of the galaxies in our sample, the spatially-extended, narrow kinematic components most likely trace primarily outflowing gas triggered by past feedback events, and the outflows have slowed down as the gas has expanded into the CGM of the host galaxy. This interpretation is based on the physical extent of the narrow components beyond the stellar continuum, the kinematics decoupled from the stars in the galaxy, the emission line profiles that are broader than the $\sigma$ obtained for the stars, and considerations regarding the SFHs of these systems. Details of individual galaxies, including the SFHs, are given below in Section~\ref{sub:sfh_link}.

Another unambiguous sign that the galaxies in our sample host outflows is provided by the \mgii \ absorption lines (Figures~\ref{fig:integrated}-\ref{fig:int_3}). The approaching sides of winds are backlit by the galaxy's stellar continuum, and thus blueshifted when observed in absorption. The HizEA galaxies commonly display extreme kinematics in the \mgii \ absorption lines \citep{tre07}. The \mgii$\lambda\lambda$2796, 2803 absorption troughs are studied in great detail in \citet{per23} using high spectral resolution Keck/HIRES data. The HIRES data show that the absorption troughs are produced by the blending of multiple kinematic components, with maximum blueshifted velocity, $v_{98}$, ranging from $-620$ to $-2700$ \kmps,  with an average value of $-1630$ \kmps. Such large line blueshifts are unambiguous signs of outflowing gas. 

Despite the high spectral resolution of the HIRES long-slit data providing robust determinations of key physical properties such as the column density, covering fraction, and mass outflow rate, such data offer only a partial view of the outflow. We find remarkable agreement between the \mgii \ absorption line profiles in the HIRES long-slit spectra and the KCWI spatially-integrated spectra, although the latter is affected by the lower spectral resolution. Interestingly, we find that \mgii \ absorption is commonly observed at large radii in the KCWI mosaics, as far as the stellar continuum extends to backlight the wind. This again reveals that the galaxies in our sample host powerful galactic-scale outflows.

\begin{figure*}[htp!]
 \centering
 \includegraphics[width=0.9\textwidth]{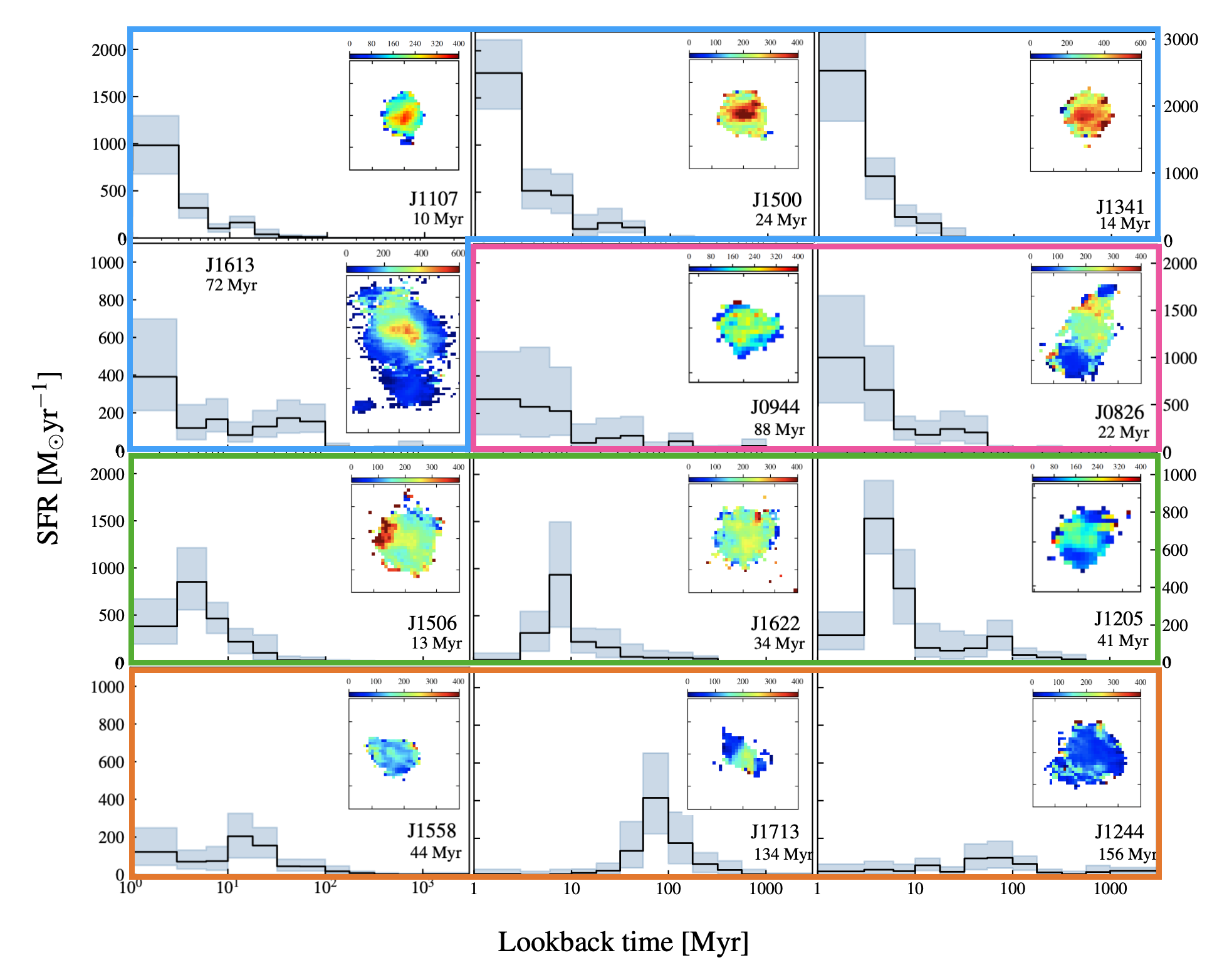}
 \caption{SFHs for the galaxies in our sample, derived with Prospector \citep{johnson21}. We mark the panels with different colors to highlight galaxies with similar SFHs showing a similar trend in the \oii \ spatial and kinematic maps (see Section~\ref{sub:sfh_link} for more details). A different scale is used for the y-axis of the top three right panels to better display the data. Light blue shaded regions show the errors on the SFH. Time is displayed on a logarithmic scale to emphasize the recent SFH. Indeed, on a linear scale, the young bursts are so impulsive they are nearly invisible against the y-axis. The mean light-weighted age of the stellar populations younger than $\sim$1 Gyr is reported at the bottom right of each panel. The \oii \ $\sigma$ map is shown for each galaxy in the upper-right inset as presented in Section~\ref{sec:kin}. The color bars show the $\sigma$ in \kmps.}
 \label{fig:sfh}
\end{figure*}

\subsection{Connection Between \oii\ Extents, \oii\ Kinematics, and Star Formation Histories \label{sub:sfh_link}}

Exploring the interplay between galactic outflows and star formation activity offers insights into the feedback mechanisms that govern a galaxy's evolution. The galaxies in our sample show extreme and ``bursty'' star formation episodes that likely drive the observed outflows, which can extend far into the CGM of the galaxy \citep{rup19}. As mentioned in Section~\ref{makani}, our team uncovered in the galaxy Makani two distinct outflow components traced by \oii \ emission whose velocities and sizes map exactly to two recent starburst episodes that this galaxy experienced 0.4 Gyr and 7 Myr ago, seen in its star formation history (SFH). To understand if the \oii \ nebulae observed here in this larger sample are also connected to the highly impulsive ``burstiness'' of the star formation in these galaxies, we investigate potential correlations between the \oii \ emission line kinematics and the SFHs of each source.

Figure~\ref{fig:sfh} shows the SFHs for the galaxies in our sample derived using Prospector \citep{johnson21}, as described in Section~\ref{subsection:properties}. The mean light-weighted age of the stellar population younger than $\sim$1 Gyr is reported in the bottom right of each panel. To ease comparison with the \oii \ kinematics, the \oii \ $\sigma$ map (as presented in Figures~\ref{fig:quad_1}$-$\ref{fig:1613_kin}) is reproduced for each target in the upper-right inset. The order of the galaxies differs from the order in earlier figures and follows the four trends described below.

Finding a simple correspondence between outflow properties traced by emission lines and the number of observed bursts in the SFH of a galaxy is challenging due to the complex geometry of the gas and projection effects. The observed emission is due to the projected signal of emitting gas filling the entire volume in front of and behind the galaxy. Projection effects, where different structures along the line of sight overlap on the plane of the sky, can alter the intrinsic width of the observed emission line and shift the central and maximum velocities ($v_{02}$ and $v_{98}$) of the velocity profile. Centrally-concentrated dust can further modify the observed line profiles.

Despite these limitations, we observe four trends when grouping the \oii \ nebulae of galaxies with similar SFHs.

1) The galaxies with the largest central velocity widths $\sigma$ (J1107, J1341, J1500, and J1613; highlighted in blue in Figure~\ref{fig:sfh}) are the galaxies with the most recent bursts. Their SFHs display a substantial increase in star formation rate in the last 3 Myr, by at least a factor of two compared to previous times. In the central regions of these galaxies, the high values of $\sigma$ of several hundreds of \kmps\ and maximum blueshifted velocities ranging between 700 and 1800 \kmps (see Table~\ref{table3}) clearly identify young outflows. Their \oii \ $\sigma$ maps show a decreasing gradient in $\sigma$ towards the outskirts of the nebulae, extending beyond the stellar continuum. This larger scale, slower gas is most likely due to an outflow driven by a previous episode of star formation that slowed as it shock-heated the surrounding gas while expanding, which would lead to narrower \oii \ emission line profiles. The J1613 nebula stands out among these four galaxies as it is more extended (even excluding the southern lobe that traces tidal interactions with a nearby galaxy) and exhibits a steeper gradient in $\sigma$ towards the outskirts. This difference may be due to the J1613 SFH having an older, prolonged burst starting $\sim$100 Myr ago as opposed to J1107, J1341, and J1500, which have no comparable bursts in the same time frame. Indeed, the J1613 light-weighted age (72 Myr) is substantially larger than the other three galaxies (10, 14, and 24 Myr). This results in the older J1613 outflow episode having more time to expand, reaching farther out in the CGM of the galaxy. 

2) Galaxies with recent star formation that is not as bursty have moderate $\sigma$ in the center of their kinematic maps. For example, the SFHs of J0826 and J0944 (highlighted in magenta in Figure~\ref{fig:sfh}) are similar to the previous four galaxies in that their star formation rate peaks in the last 3 Myr. However, their $\sigma$ maps do not exhibit comparably high and spatially concentrated values at the center. This difference may be caused by the more gradual rise of their star formation rates in the last 10 Myr compared to J1107, J1341, J1500, and J1613 (note the different y-axis scales in Figure~\ref{fig:sfh}). The outflows in J0826 and J0944 could be older and more evolved, having had more time to expand and interact with the surrounding medium. In both of these galaxies we again observe a decreasing gradient in $\sigma$ towards the outskirts of the nebulae beyond the stars, which we attribute to an older outflow episode connected to a burst of star formation seen in their SFHs around 50 Myr ago. We note that the $\sigma$ map of J0826 is one of the most complex in our sample, and the narrowest \oii \ line profiles are likely due to gas concentrated around a prominent tidal tail extending to the southeast of the galaxy.

3) Galaxies with moderate velocity dispersion and a slightly older starburst have more evolved outflows. J1205, J1506, and J1622 (highlighted in green in Figure~\ref{fig:sfh}) all have the largest burst of star formation in the last 10 Myr with a clear decrease in the most recent 3 Myr. The \oii \ $\sigma$ maps of J1506 and J1622 both show disturbed kinematics with the highest $\sigma$ values lying off-center, along an edge of each nebula. J1205 has a more symmetric $\sigma$ map, with lower $\sigma$ values towards the outskirts of the nebula, extending beyond the stellar continuum. This difference may be due to J1205 having a burst of star formation around 100 Myr ago that resulted in an outflow episode that slowed down with time as it expanded in the CGM of the galaxy.

4) Galaxies with the lowest velocity dispersion and the oldest starbursts, peaking 10--100~Myr in the past, do not have a second kinematic component. The integrated spectra of J1244, J1558, and J1713 (highlighted in orange in Figure~\ref{fig:sfh}) do not require a second Gaussian component to fit the \oii \ emission. This suggests that these galaxies did not have a recent outflow episode. J1558 has the youngest light-weighted age of these three galaxies (44 Myr) and had a burst around 30 Myr ago. J1244 and J1713 have the oldest light-weighted ages in our sample (156 and 134 Myr, respectively) and show a burst of star formation around 100 Myr ago. The absence of a substantial recent burst is likely the reason that we do not observe a broad component in the integrated spectra of these galaxies. Their $\sigma$ maps exhibit on average the lowest $\sigma$ in the sample, with values as low as $\sim$30 \kmps. J1244 and J1558 show somewhat higher $\sigma$ values, mostly off-center and beyond the tidal tail (southwest) and ongoing merger region, respectively, and are most likely due to secondary shocks. J1713 is a type II AGN candidate and the high $\sigma$ region observed at the center may be due to AGN activity.

The correspondence between the number of bursts in the SFH of a given galaxy and the kinematics of the ionized gas nebulae may not be as straightforward in the larger sample presented here as it was for Makani. Nevertheless, these four trends provide insights into the feedback mechanisms at play in our sample, where the SFH does appear to be relevant. Much of the difference between Makani, which has two clear bursts in its history and two clear outflow episodes with distinct velocities, and the rest of our sample may be due to orientation effects. Most of the outflowing gas in Makani is likely expanding in the plane of the sky and is observed as the hourglass shape of limb-brightened bipolar outflows. The outflows in the larger galaxy sample are likely propagating with different orientations compared to the line of sight and appear rounder than the Makani nebula. As a consequence, the emission line profiles in this study are likely more impacted by projection effects and therefore more difficult to interpret. We note that J1613 is similar to Makani in that the J1613 nebula has several evacuated regions, and in the central $\sim$20 kpc there is a second outflowing gas component.

The left panel of Figure~\ref{fig:3pan} compares the outflow velocity with the light-weighted age for the galaxies presented in this work and Makani. For eleven galaxies in this work we adopt the median v$_{98}$ across the entire \oii \ nebula as the outflow velocity. For the remaining galaxy, J1613, we restrict the median v$_{98}$ to the north lobe of the \oii \ nebula as the south lobe traces the interaction with a nearby by galaxy rather than outflowing gas. For Makani, we can separate the outflows into two episodes. For episode I, the oldest and slowest, we adopt the median $\sigma$ as outflow velocity as it is more representative of the extended outflow. For episode II, we use the median v$_{98}$. Unlike Makani, we are not able to distinguish two separate outflow episodes in our data, even in galaxies that show two distinct bursts in their SFH. This limitation is due to the lack of sufficient signal in individual spaxels, leading us to use a single outflow velocity in Figure~\ref{fig:3pan}. Consequently, the adopted velocity averages over the potential multiple outflows when present. We note that also the light-weighted age typically averages over two bursts, when present. Despite this limitation, a trend is evident where faster outflows (i.e. most negative velocity) are associated with galaxies that have undergone more recent bursts of star formation and possess younger stellar populations. This trend implies a link between star formation activity and the origin of these outflows. Galaxies experiencing more recent bursts of star formation are likely to have faster outflows, possibly due to the energy injection from massive young stars and the associated feedback mechanisms such as supernovae explosions and stellar winds. The scatter to lower velocity values in the figure could be attributed to projection effects which can cause the observed velocity to appear smaller than its intrinsic velocity, leading to a spread of data points towards lower velocity values. There is an absence of galaxies in the lower right of the plot, in that high-velocity outflows are not typically seen in galaxies with older stellar ages and older bursts of star formation.  These results are consistent with models in which outflows decelerate with time. In particular, the observed trend of lower outflow velocity with larger light-weighted age matches the analytic models of \citet{loc18}, where impulsive bursts of star-formation-driven winds slow down and cool as they expand into the CGM.

\subsection{Uncertainty in Mass Outflow Rate Estimation}\label{sub:mout}

To understand the potential impact of the observed outflows on the evolution of their host galaxies, it is critical to estimate the mass and energy they carry out of the galaxies. Deriving these quantities accurately poses a challenge, however, as it requires precise knowledge of the outflow geometry and kinematics, as well as the physical conditions of the gas within the wind and the properties of the ISM where the outflow propagates. While first-order estimates derived from observed emission line properties are often informative, here we choose not to report such estimates due to their high uncertainty.  We briefly discuss our reasoning below. 

H$^+$ is the most abundant species in the ionized gas phase and can be used to estimate the mass of the outflowing ionized gas ($M_{ion}$) by counting the number of recombining hydrogen atoms \citep{ost06}. In this context $M_{ion}$, is proportional to the luminosity of an H recombination line (like H$\alpha$) and inversely proportional to the electron density of the ionized gas (n$_e$). 
In the absence of spatially-resolved recombination line maps, single spectra are used to estimate $M_{ion}$.
For our sample, we can follow \citet{rup19} and use single-aperture H$_{\alpha}$ or H$_{\beta}$ measurements \citep{per21,ala15} to estimate $M_{ion}$, assuming an \oii \ to H$_{\alpha}$ ratio of 1.0 and median n$_e$ of 530 cm$^{-3}$ \citep{per21}. These assumptions yield a median $M_{ion}$ of 1.75 $\times$ 10$^8$ M$_{\odot}$. Finally, adopting R$_{90}$ (Table~\ref{table3}) as the single radius of the wind and the median v$_{98}$ within the full \oii \ nebula as its velocity yields a median value of $\dot{M}_{out}$ for our sample of 4.5 M$_{\odot}$ yr$^{-1}$ \citep[e.g.][]{leu19, rup19}. 

This estimate is highly uncertain due to systemic uncertainties in the \oii \ to H$_{\alpha}$ ratio and $n_e$. For instance, \citet{rup23} uses Keck/ESI long-slit data to show that Makani's wind has complex excitation and density structures. The \oii/H$_{\alpha}$ ratio rises from 1/3 in the center to 2 at 30$-$40 kpc, meaning a constant ratio equal to unity underestimates the mass outflow rate by about a factor of 3. They find $_e \sim 200$ cm$^{-3}$ at the center and n$_e > 2500$ cm$^{-3}$ for the blueshifted component, with much lower n$_e$ ($< 10$ cm$^{-3}$) at larger radii. Adopting a single n$_e$ for Makani would lead to errors that are orders of magnitude in size. Other studies reveal similar complexities in $n_e$ distributions, with high-density regions often confined to small areas and much lower densities observed elsewhere \citep{rup17, kak18, min19, dav20}.

A complementary approach is high-resolution absorption line studies. While emission lines probe the projected signal of gas filling the entire volume in front of and behind the galaxy, absorption lines probe only the gas along the line of sight. The kinematics of absorbing gas can be more easily separated from the underlying stellar component, especially when the absorbers are blueshifted significantly with respect to the galaxy's systemic velocity. Furthermore, absorption lines are sensitive to the density of the gas probed. This results in absorption lines providing access to lower density material that would otherwise be missed.

In \citet{per23} we present high spectral resolution (8 \kmps) optical Keck/HIRES spectra of 14 of the starburst galaxies (six of which are in the sample presented here), covering a suite of Mg and Fe absorption lines. This data allowed us to derive accurate measurements of column density, covering fraction, and robust mass outflow rates of $\dot{M}_{out}$ of $\sim 50-2200$ M$_{\odot}$ yr$^{-1}$. These values are up to {\it three orders of magnitude} larger than suggested by the simple estimate above for the ionized outflow. The $\dot{M}_{out}$ values derived in \citet{per23} are lower limits for most of the galaxies, as we have bounds on the optical depth from unsaturated \ion{Fe}{2} absorption lines for only 36\% of the detected \ion{Mg}{2} absorption troughs. Although these estimates are lower limits, they are more reliable than emission line-derived values and suggest that these starburst galaxies are capable of ejecting very large amounts of cool gas that will substantially impact their future evolution.

\subsection{Implications for Starburst-driven Feedback}\label{sub:implications}

Our sample provides an excellent laboratory to test the limits of stellar feedback and evaluate whether stellar processes alone can drive powerful outflows, without the need to invoke undetected AGN feedback.
In some models, AGN feedback occurs at a critical stage in galaxy evolution during a major merger or accretion event which then triggers a massive burst of star formation and rapid accretion onto the central SMBH. Observationally, starburst galaxies are typically shrouded in gas and dust \citep[e.g.][]{san88, hop05, hop08, vei09}. AGN feedback occurs when energy and momentum released by the accreting SMBH couples to the surrounding ISM, leading to a powerful `blowout' of gas and dust that can then help quench star formation, potentially revealing a visibly luminous quasar in the galactic nucleus \citep[e.g.][]{san88, dim05, hop06, hop08, hop16, rup11, rup13, liu13}. An important question for the HizEA galaxy sample presented here is understanding how it fits into this merger-driven galaxy evolution scenario. 

The HizEA galaxies are late-stage major mergers with exceptionally compact central star-forming regions, and a substantial portion of their gas and dust is blown away by powerful outflows \citep{dia12, sel14, per23, rup23}. Their dense, dusty cores show a complex “picket fence” ISM geometry with some high attenuation sightlines and some holes. The high incidence of powerful outflows detected may be responsible for such holes in the ISM \citep{per21}. Most galaxies in our sample exhibit multiple outflow episodes, which we show here are connected with their SFH, and their spatial extent traced by emission lines extends beyond the stellar continuum. We find little evidence of ongoing AGN activity in these systems based on X-ray, infrared, radio, and spectral line diagnostics \citep{sel14, per21, rup23}. For the galaxies that may contain a dust-obscured accreting SMBH, the AGN is not currently energetically dominant \citep{dia12, sel14} or radio-loud \citep{pet20}.  While we cannot rule out past heightened AGN activity, multi-wavelength data for the vast majority of these galaxies can be explained by their known extreme star formation properties and the possible presence of shocks \citep{rup19, per21, per23, rup23}. These results agree with models in which stellar processes can drive extreme outflows, without requiring feedback from SMBHs \citep[e.g.,][]{hop12, hop14, dru18, pan21}.

Further observational evidence supporting this evolutionary picture for the HizEA galaxies comes from low ionization absorption lines tracing the cool, ionized outflow phase. Studies characterizing the gas in proximity to luminous quasars have found a considerably different ionization state between gas across (transverse direction) and along the line of sight (down the barrel) to the quasar \citep[e.g.][]{pro14, per16}. In particular, there is a deficiency of cool gas, as traced by low ions like \mgii \ absorbers, in down the barrel observations of quasars, in contrast to studies based on projected quasar pairs which report a high \mgii \ incidence extending to 200$-$300 kpc in the transverse direction of the quasar \citep{far14, pro14, per16, per18, lau18}. This finding indicates that the ionizing emission from quasars is highly anisotropic, and one should not expect to observe strong \mgii \ absorption along the propagation axis of an outflow driven by a quasar due to the high ionization field.

As mentioned in Section~\ref{sub:outflows}, we observe strong \mgii \ absorption lines in all of the KCWI datacubes presented here, to the full extent that the stellar continuum exists to provide a background light source against which to see the absorption. The \mgii \ absorption troughs have maximum velocities of 820--2860 \kmps \ \citep{dav19, per23} and probe powerful galactic-scale outflows that are propagating along the line of sight, though we do not have an exact knowledge of their geometry and orientation due to projection effects. The presence of strong \mgii \ absorbers along the line of sight to the HizEA galaxies stands in contrast to the hypothesis that the
observed outflows may represent a `blowout' phase powered by recent quasar activity, unless enough time has passed for all the observed \mgii \ to recombine (though this seems unlikely). It is therefore likely that these outflows are driven by stellar processes alone.

\begin{figure*}[htp!]
 \centering
 \includegraphics[width=\textwidth]{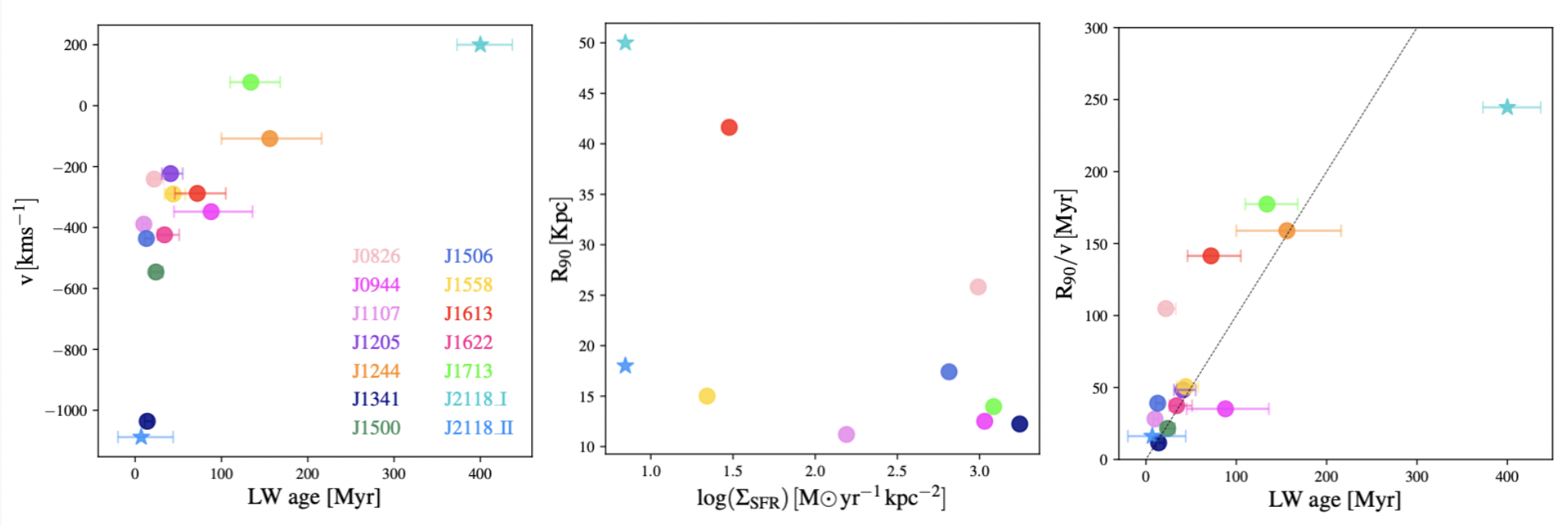}
 \caption{Left: Outflow velocity, v, vs. light-weighted age. Center: \oii \ radius enclosing 90\% of the surface brightness, R$_{90}$ vs. $\rm \Sigma_{SFR}$. Right: Outflow dynamical time vs. light-weighted age. The filled circles represent the galaxies studied in this work and the filled star is the galaxy Makani (see legend in the first panel). The tilted line in the central panel illustrates outflow timescales that match 1:1 the light-weighted age.}

 \label{fig:3pan}
\end{figure*}

The central panel of Figure~\ref{fig:3pan} compares the size of the \oii \ nebulae as represented by the radius enclosing 90 percent of the surface brightness, R$_{90}$, and the galaxy star formation rate surface density, $\Sigma_{SFR}$. We note that the galaxies with the highest $\Sigma_{SFR}$ have smaller outflows, implying they were created more recently. The presence of smaller outflows in galaxies with high $\Sigma_{SFR}$ implies that the intense star formation activity in these galaxies is providing the energy necessary to drive the observed outflows. The absence of galaxies with high $\Sigma_{SFR}$ and large sizes can be interpreted in the context of the timescales involved in the evolution of the outflows, as it takes time for the outflows to propagate to large scales. 

To directly compare the outflow and star formation timescales, we plot in the right panel of Figure~\ref{fig:3pan} the outflow dynamical time $t_{out}\sim$ (R$_{90}$/v) versus the light-weighted age. Despite likely projection effects, in many systems these timescales match to within a factor $\sim$2. In outflows with older light-weighted ages and larger nebulae, such as in Makani and J1613, the larger-radius gas has decelerated but not entirely dispersed, and this timescale is primarily applicable to the fast, recent outflow. (E.g., we estimate a dynamical timescale for the early, Episode~I wind in Makani of 400~Myr.) The calculated outflow timescales are $t_{out} \la 50$~Myr. Thus, the outflows are expanding and mixing with the CGM on timescales $\ga$50~Myr.

The compact starburst galaxies in the HizEA sample could represent a short but relatively common phase of massive galaxy evolution \citep{wha22}.  They are characterized by multiple bursts of star formation triggered by mergers. While we observe these galaxies during an extreme phase of their evolution, it is possible that most massive galaxies experience such merger-induced bursts of star formation which drive powerful outflows that may play a critical role in quenching their star formation.

Furthermore, the HizEA sample provides potentially powerful constraints on the physics of multiphase galactic winds driven by star formation feedback. Recent advances on the theory of such outflows paint a picture consistent with our observations and may help explain the overall energetics and mass fluxes of such winds, as well as the acceleration mechanisms for the cool gas traced by \oii \ \citep[e.g.,][]{Fielding:2022, Kim:2020, Schneider:2024}. The extreme nature of the HizEA sample can shed light on the limits and accuracy of current state-of-the-art analytic and simulation-based theoretical models.

\section{Summary and Conclusions}\label{sec:conclusion}

We use new optical Keck/KCWI IFU data of 12 massive, compact starburst galaxies at $z \sim 0.5$ to probe the
structure of their extreme, ejective feedback episodes and investigate the potential impact of these outflows on the evolution of their host galaxies. These galaxies are massive ($\rm M_* \sim$10$^{11} \, M_{\odot}$), extremely compact (half-light radius $\sim$few hundred pc), have very high star formation rates (mean $\rm SFR \sim200 \, M_{\odot} \ yr^{-1}$) and star formation surface densities (mean $\rm \Sigma_{SFR} \sim2000 \, M_{\odot} \ yr^{-1} \ kpc^{-2}$), and are known to drive extremely fast (maximum velocities of 820$-$2860 \kmps) outflows traced by \mgii \ absorption lines \citep{tre07, dav19, per23}. The KCWI data cover \oii$\lambda\lambda 3726,3729$ \ and \mgii$\lambda\lambda 2796,2803$ \ and allow us to directly measure the morphology, physical extent, and resolved kinematics of the outflows' cool ($T\sim 10^4$ K), ionized gas phase as traced in emission.

Our main conclusions are:
\begin{itemize}
\item  We detect \oii \ emission nebulae in all 12 galaxies spanning a wide range of shapes and physical extents (Section~\ref{sec:mor}), with projected areas of 500 to 3300 kpc$^2$ and maximum radial extents of 10 and 40 kpc (Figure~\ref{fig:o2_sb} and Table~\ref{table3}). These are among the largest \oii \ nebulae observed around isolated  galaxies.

\item  In all galaxies the \oii \ emission is spatially extended beyond the stellar continuum. Despite the diversity of morphologies, sizes, and luminosities, the \oii \ surface brightness profiles are similar. They are shallow in the central regions and gradually decrease with radial distance, less steeply than the stars (Section~\ref{sec:rad_profile} and Figure~\ref{fig:sb_profile}).

\item  In these galaxies, \oii \ is a more effective tracer than \mgii \ for detecting low surface brightness extended emission, with \mgii \ consistently showing a weaker signal than \oii. The observed \mgii \ emission is substantially less extended than the \oii \ emission (Section~\ref{sec:eml} and Figure~\ref{fig:mg2_sb}).

\item  The \oii \ nebulae in our sample are dominated by non-gravitational motions and trace galactic outflows with maximum blueshifted speeds, $v_{98}$, ranging from $-$335 to $-$1920 \kmps. This is evident from their physical extension beyond the stellar continuum and their kinematics, which deviate from dynamical equilibrium with the host galaxy or disk rotation (Section~\ref{sub:outflows} and Figure~\ref{fig:kin_gas_star}). 

\item  We find clear observational trends when grouping the \oii \ properties of galaxies with similar star formation histories, suggesting a correlation between the galactic outflows observed and the intense and bursty star formation in our sample (Section~\ref{sub:mout} and Figure~\ref{fig:sfh}). These trends include: galaxies with recent intense star formation show high velocity dispersions at their centers, indicative of young outflows; galaxies with older bursts of star formation have larger, more evolved outflows, with lower velocity dispersions; and galaxies with the oldest starbursts lack recent outflows.

\item Mass outflow rates derived from high-resolution absorption line data for these galaxies, presented in \citet{per23}, ($50 - 2200$ M$_{\odot}$ yr$^{-1}$) are more reliable---and up to three orders of magnitude larger---than those derived from emission line properties in this study (Section~\ref{sub:mout}). Estimating mass outflow rates from the current emission line data is uncertain due to unknowns in gas geometry, ionization state, and density. 

\item Observations do not support on-going AGN activity in these galaxies  (Section~\ref{sub:implications}) and instead align with models where powerful outflows are primarily driven by extreme stellar feedback.
    
\end{itemize}

The galaxy sample studied here provides a unique opportunity to study star formation and feedback at its most extreme. In a forthcoming paper based on Keck KCRM spectra, we will determine more accurate physical properties of these galactic winds thanks to a full suite of optical emission lines. Such studies crucially inform theorists working to understand the mass and energy content in outflows. These new data will directly test the latest models of how outflows populate the CGM of massive galaxies with cool gas, a major open question in both CGM and galactic outflow science.

\section*{Acknowledgements}

We thank the referee for providing a constructive report that has helped to improve the clarity of the manuscript.
We acknowledge support from the Heising-Simons Foundation grant 2019-1659. S.~P. and A.~L.~C. acknowledge support from the Ingrid and Joseph W. Hibben endowed chair at UC San Diego. The data presented herein were obtained at the W. M. Keck Observatory, which is operated as a scientific partnership among the California Institute of Technology, the University of California, and the National Aeronautics and Space Administration. The Observatory was made possible by the generous financial support of the W. M. Keck Foundation.
The authors wish to recognize and acknowledge the very significant cultural role and reverence that the summit of Maunakea has always had within the indigenous Hawaiian community. We are most fortunate to have the opportunity to conduct observations from this mountain.

\bibliography{biblio}{}
\bibliographystyle{aasjournal}

\appendix
\section{Fits of Individual Targets}\label{app:A}
\restartappendixnumbering

\begin{deluxetable*}{crcrclcrrrrc}
\tablecaption{Integrated \oii \ Kinematics\label{table6}}
\tablehead{
\multicolumn{1}{c}{} &
\multicolumn{4}{c}{Nebula} &
\multicolumn{1}{c}{} &
\multicolumn{1}{c}{} &
\multicolumn{4}{c}{Core} \\
\cline{2-6} \cline{8-12}
\multicolumn{1}{c}{} &
\multicolumn{2}{c}{Narrow} &
\multicolumn{2}{c}{Broad} &
\multicolumn{1}{c}{B/N} &
\multicolumn{1}{c}{} &
\multicolumn{2}{c}{Narrow} &
\multicolumn{2}{c}{Broad} &
\multicolumn{1}{c}{B/N} \\
\multicolumn{1}{c}{Object Name} &
\multicolumn{1}{c}{$v_{50}$} &
\multicolumn{1}{c}{$\sigma$} &
\multicolumn{1}{c}{$v_{50}$} &
\multicolumn{1}{c}{$\sigma$} &
\multicolumn{1}{c}{flux} &
\multicolumn{1}{c}{} &
\multicolumn{1}{c}{$v_{50}$} &
\multicolumn{1}{c}{$\sigma$} &
\multicolumn{1}{c}{$v_{50}$} &
\multicolumn{1}{c}{$\sigma$} &
\multicolumn{1}{c}{flux} \\
\colhead{(1)} & \colhead{(2)} & \colhead{(3)} & \colhead{(4)} & \colhead{(5)} & \colhead{(6)} & & \colhead{(7)} & \colhead{(8)} & \colhead{(9)} & \colhead{(10)} & \colhead{(11)}
}
\startdata
J0826+4305 & $-$19 & 171 & $-$69 & 490 & 0.82 & & $-$24 & 183 & $-$155 & 712 & 0.26 \\
J0944+0930 & $-$4 & 102 & $-$92 & 279 & 4.4 & & 25 & 80 & $-$140 & 257 & 2.27 \\
J1107+0417 & 3 & 201 & $-$65 & 690 & 0.88 & & 20 & 212 & $-$38 & 671 & 0.50 \\
J1205+1818 & $-$5 & 176 & $-$390 & 450 & 0.19 & & $-$18 & 167 & $-$340 & 452 & 0.16 \\
J1244+4140 & 56 & 152 & \nodata & \nodata & \nodata & & 61 & 148 & \nodata & \nodata & \nodata \\
J1341$-$0321 & $-$90 & 202 & $-$463 & 564 & 4.47 & & $-$59 & 325 & $-$787 & 398 & 0.88 \\
J1500+1739 & $-$20 & 282 & $-$7 & 628 & 0.79 & & $-$68 & 251 & 19 & 1029 & 0.55 \\
J1506+5402 & $-$57 & 216 & $-$272 & 825 & 0.61 & & $-$60 & 216 & $-$230 & 726 & 0.27 \\
J1558+3957 & $-$96 & 184 & \nodata & \nodata & \nodata & & $-$116 & 190 & \nodata & \nodata & \nodata \\
J1613+2834 & 17 & 229 & $-$226 & 584 & 0.79 & & $-$49 & 259 & $-$335 & 713 & 0.31 \\
J1622+3145 & $-$60 & 219 & $-$366 & 465 & 0.40 & & $-$110 & 229 & $-$110 & 229 & 0.25 \\
J1713+2817 & 39 & 396 & \nodata & \nodata & \nodata & & 90 & 188 & \nodata & \nodata & \nodata \\
\enddata
\tablecomments{Col 2 - 5: narrow and broad components central velocity ($v_{50}$) and line width ($\sigma$) of the \oii \ line profile integrated over the whole nebula; Col 6: Flux ratio of the Broad to Narrow components of the \oii \ line profile integrated over the whole nebula; Col 7 - 10: narrow and broad components $v_{50}$ and $\sigma$ of the \oii \ line profile integrated over the core of the nebula (central 5 $\times$ 5 spaxels); Col 11: Flux ratio of the Broad to Narrow components of the \oii \ line profile integrated over the core of the nebula.}
\end{deluxetable*}

\begin{figure*}[htp!]
 \centering
 \includegraphics[width=0.9\textwidth]{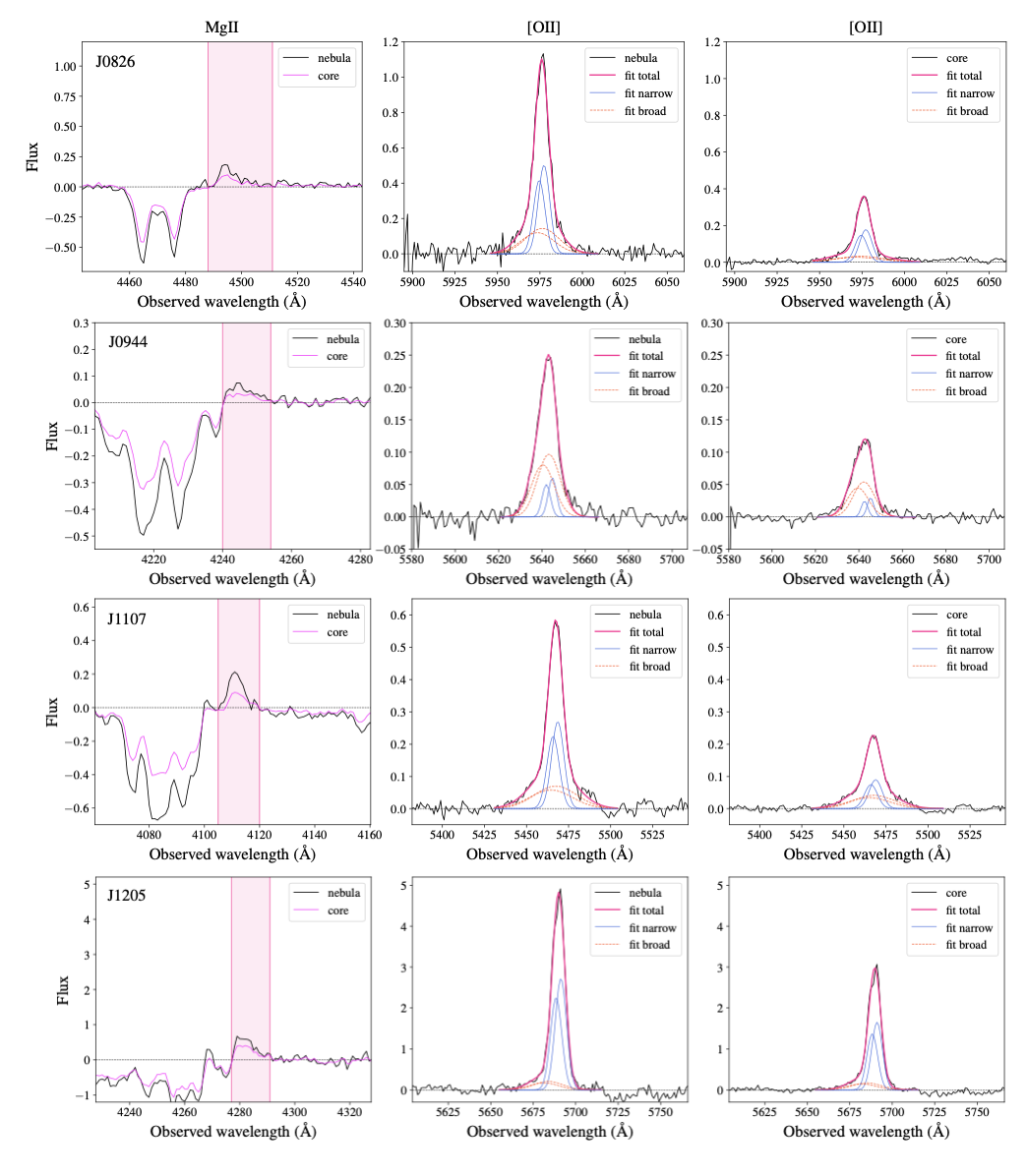}
 \caption{\mgii \ and \oii \ spatially-integrated continuum-subtracted spectra for each galaxy in our sample. Left: Weak \mgii \ emission and strong blueshifted absorption are observed in each galaxy. The solid black line shows the integrated spectrum over the full extent of the nebula, and the purple solid line shows the integrated spectrum over the nebula's core, i.e. the central 5 $\times$ 5 central spaxels. The vertical pink band represents the spectral region used to integrate the continuum-subtracted spectrum in each spaxel used to construct the  \mgii \ surface brightness map. 
 Center: Strong \oii \ emission is observed in each galaxy. The solid black line shows the spectrum integrated over the full extent of the nebula, and the magenta solid line shows the total Gaussian fit to the \oii \ emission doublet. Blue and orange lines show the fits to the individual \oii \ doublet lines for the narrow and broad components, respectively.
 Right: Same as the central figures, where here the data and fits are shown for the spectrum integrated over the core of the nebula, i.e. the central 5 $\times$ 5 central spaxels.
} \label{fig:integrated}
\end{figure*}

\begin{figure*}[htp!]

 \centering
 \includegraphics[width=0.9\textwidth]{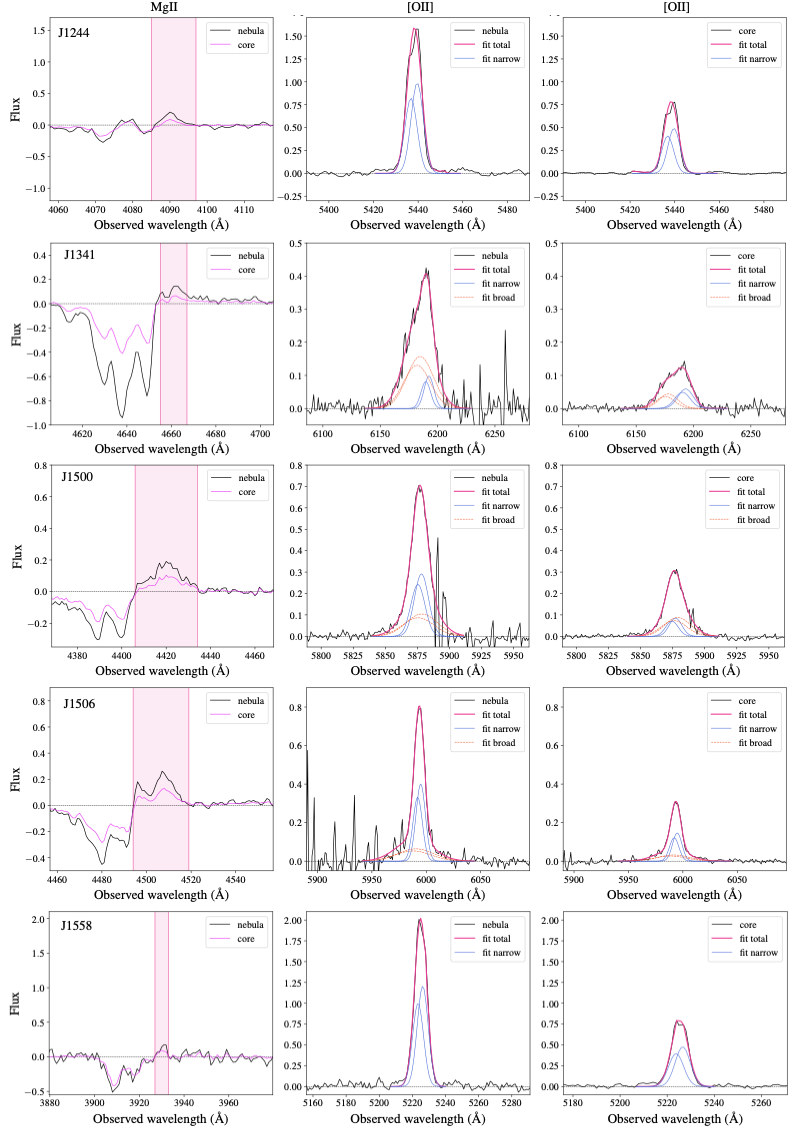}
 \caption{ -- Same as in Figure~\ref{fig:integrated}}
 \label{fig:int_2}
\end{figure*}

\begin{figure*}[htp!]
 \centering
 \includegraphics[width=0.9\textwidth]{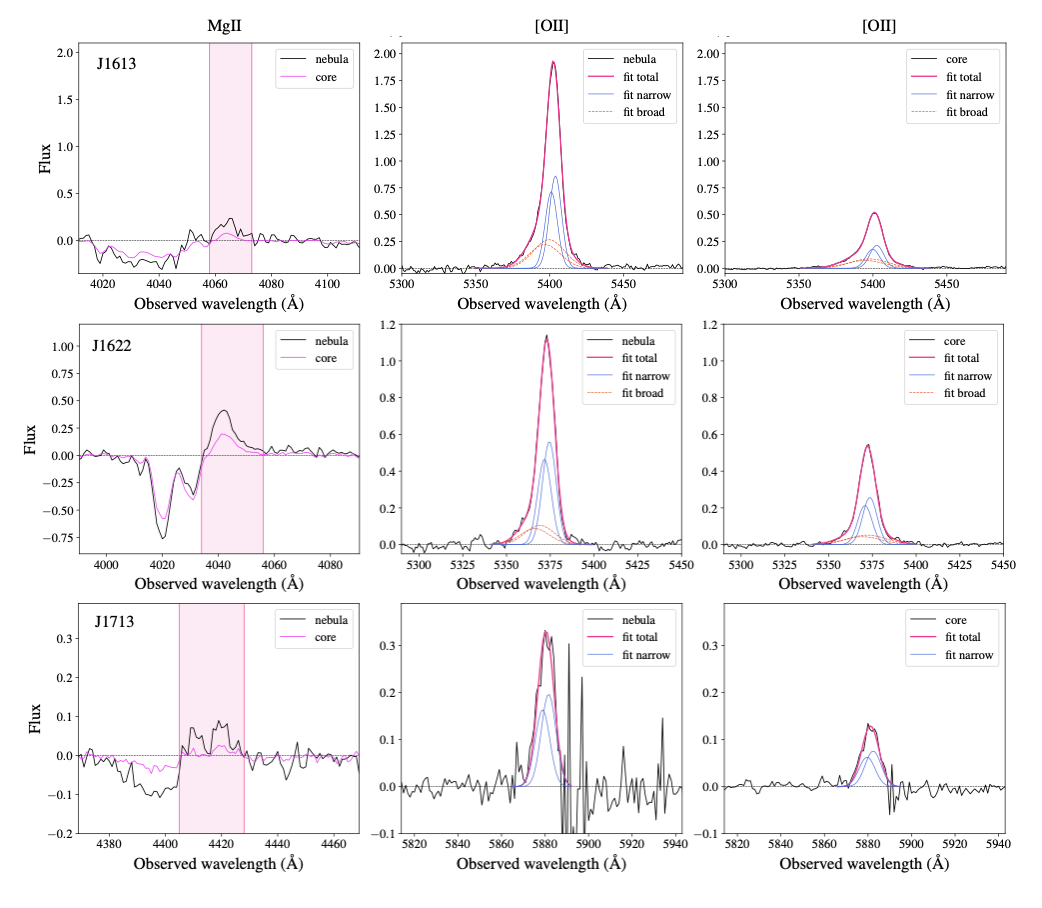}
 \caption{-- Same as in Figure~\ref{fig:integrated}}
 \label{fig:int_3}
\end{figure*}

\begin{figure*}[htp!]
 \centering
 \includegraphics[width=0.9\textwidth]{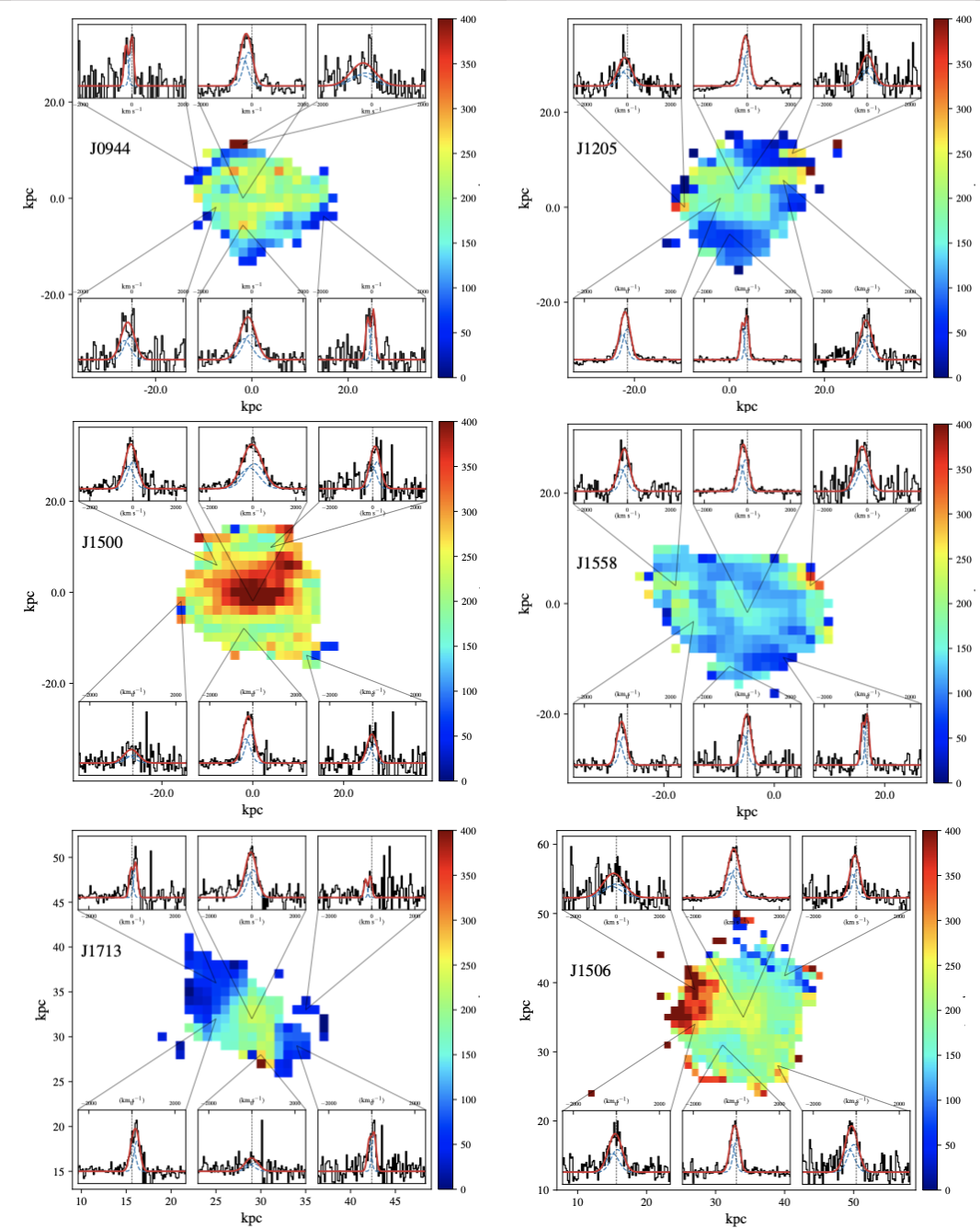}
 \caption{Velocity dispersion $\sigma$ map of the \oii \ emission for six galaxies in our sample J0944, J1205, J1500, J1558, J1713, and J1506. Velocity profiles of representative spaxels are shown in the insets, highlighting areas with high and low velocity dispersion. In the velocity profiles and spectra, the black line is the continuum-subtracted spectrum, the red solid line is the total emission line model, and the blue dashed lines are the \oii \ emission line models for the individual \oii \ doublet components.}
 \label{fig:inset_3}
\end{figure*}

\end{document}